\documentclass[twocolumn,twocolappendix]{aastex631}
\usepackage{aas_macros}
\usepackage{amsmath}
\usepackage{amssymb}
\usepackage{amsfonts}
\usepackage{booktabs, array}
\usepackage{tabularx}
\usepackage{multirow}
\usepackage{natbib}
\usepackage{tikz}
\usepackage{graphicx} 
\usepackage[]{xcolor}
\begin{document}

\title{Micro-Tidal Disruption Events at Galactic Centers}

\author[0000-0003-0750-3543]{Xinyu Li}
\affiliation{Department of Astronomy, Tsinghua University, Beijing, 100084, China}

\author{Houyi Sun}
\affiliation{Department of Astronomy, Tsinghua University, Beijing, 100084, China}

\author{Yuan-Chuan Zou}
\affiliation{School of Physics, Huazhong University of Science and Technology, Wuhan 430074, China}

\author{Huan Yang}
\affiliation{Department of Astronomy, Tsinghua University, Beijing, 100084, China}
\email{hyangdoa@tsinghua.edu.cn}

\begin{abstract}
This work explores a scenario for micro-tidal disruption events (TDEs) triggered by close encounters between high-speed white dwarfs (WDs) and stellar-mass black holes (sBHs) in galactic centers. In this model, a WD orbiting the central massive black hole (MBH) is scattered by an sBH during the sBH's early extreme mass-ratio inspiral phase. We conservatively estimate these events occur a few times per year within $z\leq 3$.
Significant disruption of the WD occurs when the impact parameter is comparable to the WD's radius. We derive a mathematical criterion and confirm numerically by hydrodynamical simulations.
With the increase of the impact parameter and the collision speed, the WD material captured by the sBH decreases while the material remain self-gravitating increases.
A part of the WD material becomes unbound from the sBH-WD system, and its mass ranges from nearly zero to $\ge 50\%$, reaching the peak value when the impact parameter is comparable to the WD's radius. We expect the subsequent capture of WD material by the sBH to produce a prompt X-ray burst (a micro-TDE), and the accretion of unbound debris onto the MBH can power a fainter, delayed optical flare. The properties of certain transient X-ray bursts observed by Einstein Probe are consistent with this micro-TDE picture.
\end{abstract}

\section{Introduction}

Recent observations of Quasi-Periodic Eruptions (hereafter QPEs) have provided evidence for the existence of stellar-mass objects orbiting massive black holes (hereafter MBH) in galactic centers at a few hundred gravitational radii \citep{Linial:2023nqs,Zhou:2024bjt}. According to the detection rate of QPEs, if they all originate from the star-disk collision scenario, it is argued that an MBH generally has one or more such neighbors \citep{Arcodia:2024efe}.

Stars are the preferred stellar-mass objects for generating QPEs, as they generally have a large impact radius with the disk. For stellar-mass black holes, to increase the Bondi radius for gas capture, the inclination of the orbit has to be small relative to the disk plane.
The underlying mechanism that populates stellar-mass objects at a few hundred gravitational radii around massive black holes is still unknown, but it is possibly associated with disk migration during the previous active phase of Active Galactic Nuclei (hereafter AGN) \citep{Pan:2021ksp,Pan:2021lyw,Pan:2021xhv}. 
The study in \citet{Pan:2021ksp} showed a local maximum at a few hundred gravitational radii for the lifetime of these stellar mass objects within a thin disk, where for larger radii the disk migration dominates and for smaller radii the gravitational wave decay dominates. 
Although stars may contribute to the majority of QPEs, the same mechanism that brings them near MBHs may also deliver stellar-mass black holes (hereafter sBHs), neutron stars, and white dwarfs (hereafter WDs). In
this work, we focus on systems with white dwarfs around a few hundreds of the gravitational radii from massive black holes at the galactic center.

\begin{figure}[h]
    \centering    \includegraphics[width=\linewidth]{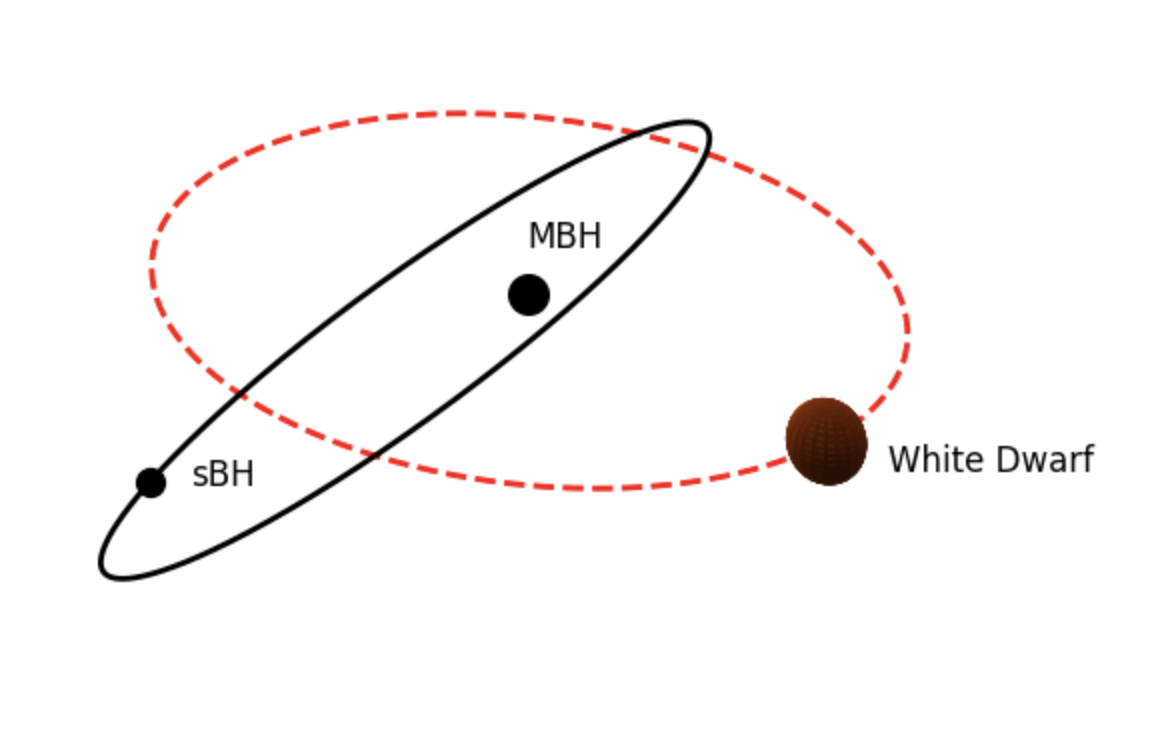}
    \caption{Schematic diagram of the orbits of a WD and a sBH, orbiting around a MBH at the galactic center.}
    \label{fig:sch}
\end{figure}

In addition to these tightly bound WDs, extreme mass-ratio inspirals (hereafter EMRIs) may form due to scattering processes in the nuclear star cluster (``dry EMRIs") and/or disk-assisted migration during the AGN active episode (``wet EMRIs"). 
In \citet{Pan:2021ksp}, it is shown that disk-assisted migration can greatly accelerate the rate of EMRI formation, so that overall wet EMRIs may be more common than dry EMRIs \citep{Pan:2021oob}. 
Focusing on the quiescent phase of the AGNs, the dry EMRI event roughly happens once every few million years for million solar mass massive black holes \citep{Pan:2021lyw}. In the early stage of evolution of these dry EMRIs, the orbit is highly eccentric. The sBH of these dry EMRIs can approach very closely (with impact parameter one or two times the WD radius) to the WD (see Fig.~\ref{fig:sch}).

During a close encounter event between a WD and a sBH, in contrast with star collisions in the nuclear star cluster \citep{Rose:2023}, the initial relative speed should be $\sim 0.1c$ according to the distance of the WD from the MBH. This contrasts with other micro-TDE scenarios considered in previous literature (often associated with stars and planets), where the initial relative speed is much smaller \citep{Xin:2023fqa,Perets:2016pwr,Kremer:2020cne}. The electromagnetic signatures are vastly different because of the difference in disruption timescale and/or mass accreted. 

In this article, we conduct a comprehensive study of the scenario of close encounter between a high-speed WD and a sBH.
In Sec.~\ref{sec:rate} we present a rate calculation for this type of close encounter events between a WD and a sBH. In Sec.~\ref{sec:theory}, we derive the criterion for whether the WD can be tidally disrupted during the encounter. In Sec.~\ref{sec:hydro} we perform a series of hydrodynamical simulations of the close encounter between a WD and a sBH with various initial impact parameter $b$ and relative speed $v_0$, and quantify the dependence of the accreted material on the sBH and the unbound material on $b, v_0$. In Sec.~\ref{sec:em} we discuss possible electromagnetic and gravitational wave signatures of this type of transient events. We conclude in Sec.~\ref{sec:con}. 

\section{Event Rate}
\label{sec:rate}

Consider a WD (with mass $m_{\rm WD}$ and radius $R_{\rm WD}$) following a circular orbit of radius $r_{\rm orbit}$ around a MBH (with mass $M$ and gravitational radius $r_g=GM/c^2$). A sBH is following a highly eccentric orbit with peri-center distance smaller than $r_{\rm orbit}$ and semi-major axis larger than $r_{\rm orbit}$. To the lowest order approximation, the sBH crosses the sphere defined by the orbit of the WD twice per orbit, and the relative orientation between the orbits is assumed to be random due to randomized initial distribution and orbital precession. Collective mechanisms such as orbital resonances between orbits are not considered. As a result, within one orbit, the sBH has approximately the probability
\begin{eqnarray}\label{eq:prob}
\mu(M) &\sim& 2 \frac{ \pi R^2_{\rm WD}}{4 \pi r_{\rm orbit}^2} \nonumber\\
&\sim& 10^{-9} \left(\frac{r_{\rm orbit}}{100 r_g}\right)^{-2} \left(\frac{M}{10^6 M_\odot}\right)^{-2}\left(\frac{R_{\rm WD}}{6000 {\rm km}}\right)^2
\end{eqnarray}
to have a close encounter with the WD where the impact parameter is comparable to the WD radius.

To compute the close encounter event rate, we adopt the same formalism used in \citet{Pan:2021ksp}.
The initial nuclear cluster model is assumed to be the Tremaine’s cluster model \citep{Tremaine:1994} with two components in the stellar cluster: stars with $m_{\rm star}=1M_\odot$ and total stellar mass $M_{\rm star} = 20M$ and sBHs with $m_{\rm sBH}=10M_\odot$ each and total mass $M_{\rm sBH}$.
The number densities in position space for stars $n_{\rm star}$ and sBHs $n_{\rm sBH}$ are given by
\begin{align}
    n_{\rm star}(r) &= \frac{M_{\rm star}}{m_{\rm star}} \frac{3-\gamma}{4\pi} \frac{r_a}{r^{\gamma}(r+r_a)^{4-\gamma}}, \\
    n_{\rm sBH} &= \delta \times n_{\rm star}(r)\,.
\end{align}
$\delta$ is the abundance of sBH relative to the stars, which is set as $10^{-3}$ in our model.
$r_a=4GM/\sigma^2$ is the density transition radius.
$\sigma$ is the stellar velocity dispersion following 
$M-\sigma$ relation \citep{Gultekin:2009} 
\begin{equation}
    M = 1.53\times10^6M_{\odot}\left(\frac{\sigma}{70\text{km/s}}\right)^{4.24}.
\end{equation}
$\gamma$ is the density scaling power index.
Mass segregation effect in initial conditions is neglected \citep{BahcallWolf:1977,Takahashi1995,Murphy2011,Pavlik2019}. It may further increase the calculated event rate. Mass segregation during the evolution of the cluster is naturally accounted for in the Fokker-Planck scheme below.
The density profiles leads to a gravitational potential of form
\begin{equation}
    \phi(r) = \frac{M}{r}+\frac{M_{\rm star}+M_{\rm sBH}}{r_a}\frac{1}{2-\gamma}\left[1-\left(\frac{r}{r+r_a}\right)^{2-\gamma}\right].
\end{equation}

The evolution of stellar-mass black holes within the nuclear star cluster can be described using the phase-space distribution function $f_i$ ($i=\rm star,\,sBH)$ of each component.
It is more convenient to parametrize the phase space using $(E,R)$ coordinate \citep{Cohn1978, Cohn1979}. 
$E=\phi-v^2/2$ is the binding energy per unit mass with $\phi$ being the gravitational potential.
$R\equiv J^2/J_c^2(E)$ is the dimensionless angular momentum variable, with $J$ being orbital angular momentum per unit mass and $J_c$ the maximum allowed value $J$ at given $E$.

The time-evolution of $f$ is governed by the orbit-averaged Fokker-Planck equation \footnote{For simplicity, subscript $i$ is omitted in the equations.} \citep{Pan:2021ksp,Pan:2021lyw,Pan:2021xhv}
\begin{align}\label{eq:ft}
    \mathcal{C}\frac{\partial f}{\partial t} = -\frac{\partial F_E}{\partial E} - \frac{\partial F_R}{\partial R}
\end{align}
The weight function $\mathcal{C}$ is defined as
\begin{equation}
    \mathcal{C} = 4\pi^2P(E,R)J^2_c(E) 
\end{equation}
with $P(E,R)$ being the orbital period.
$F_E$ and $F_R$ are flux components in $E,R$ directions in the phase space. $F_E$ is the flux passing through unit $R$ and $F_R$ passing through unit $E$.
The are given by
\begin{align}
    -F_E &= D_{EE}\frac{\partial f}{\partial E} + D_{ER}\frac{\partial f}{\partial R}+D_E f \\
    -F_R &= D_{RR}\frac{\partial f}{\partial R}+D_{ER}\frac{\partial f}{\partial E}+D_Rf\,,
\end{align}
the diffusion coefficients \{$D_{EE}, D_{ER}, D_{RR}$\} and the advection coefficients \{$D_E, D_R$\} are functions of $f(t, E, R)$, the expressions are in the Appendix A of \citet{Pan:2021ksp}.

Assuming the initial distribution function $f_i(t = 0, E, R)$ is isotropic which only depends on $E$, $f_i$ is related to $n_i$ through \citep{Tremaine:1994}
\begin{equation}
    f_i(t=0,E,R)=\frac{\sqrt{2}}{(2\pi)^2}\frac{d}{dE}\int\limits_0^E\frac{dn_i}{d\phi}\frac{d\phi}{\sqrt{E-\phi}}.
\end{equation}

We numerically evolve the Fokker-Planck equation from this initial distribution function with boundary conditions outlined in \citet{Pan:2021ksp}.The evolution time $t$ differs among nuclear clusters of different MBH masses, depending on whether they have reached a steady state. The maximum evolution timescale is the age of the Universe, $\sim 13\,\mathrm{Gyr}$.
The rate of peri-center passage $\Gamma$, for those orbits that cross $100r_g$ each orbit, can be estimated from $f$ as
\begin{equation}
\Gamma = \int\limits_{0}^{E_c} dE \int\limits_{R_l}^{R_p} dR \, \mathcal{C}\frac{f_{sBH}}{P},
\end{equation}
where $R_l=J_{lc}^2/J_c^2$ with $J_{lc} = 4 G M/c$ being the angular momentum per unit mass at loss cone boundary, $R_p$ is the value of $R$ for orbits with peri-center distance of $100r_g$, and $E_c$ is the value of $E$ for orbits with semi-major axis of $100 r_g$.
We define an averaged rate as
\begin{equation}
\overline{\Gamma} =\frac{1}{T} \int^T_0 dt\, \Gamma\,.
\end{equation}

\begin{figure}[h]
    \centering
    \includegraphics[width=\linewidth]{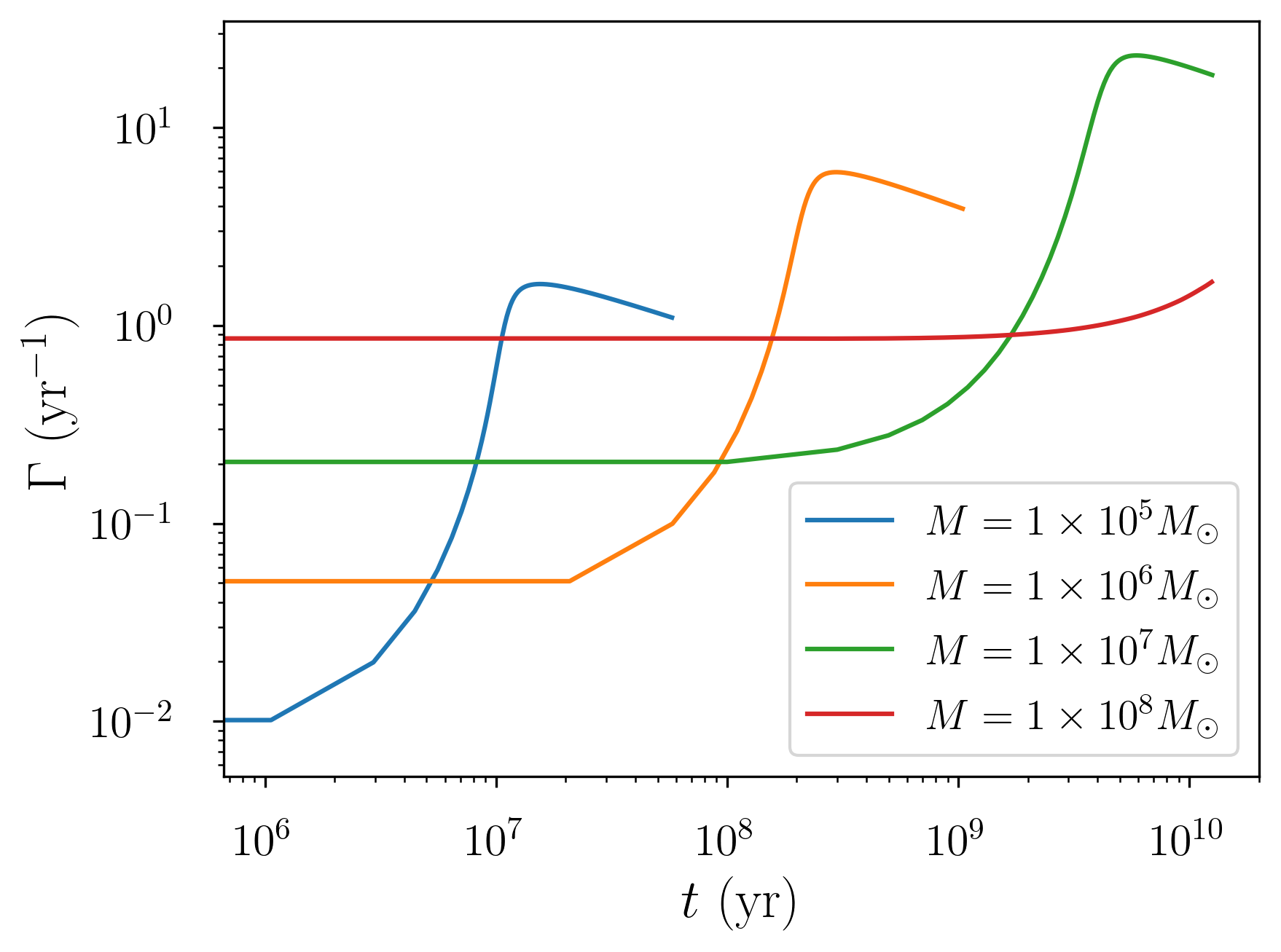}
        \label{RM}
    \hfill 
        \includegraphics[width=\linewidth]{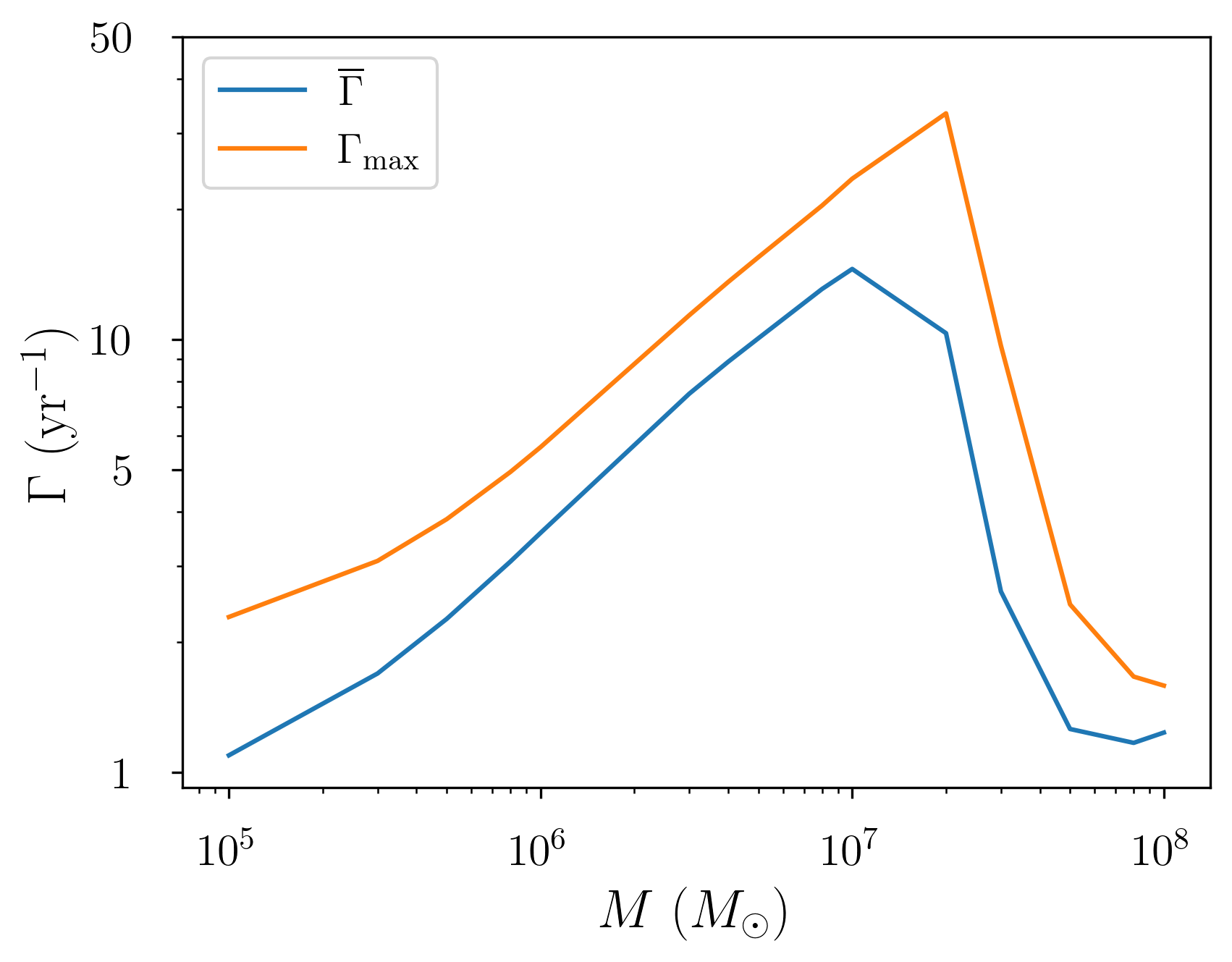}
        \label{meanRM}
    \caption{Upper panel shows evolution of the peri-center passage rate with time for the different nuclear cluster with different MBH masses. The $\Gamma$ in lower MBH mass systems ($M\lesssim10^7M_\odot$) will reach a maximum value $\Gamma_{\rm max}$, the stable state, because of the mass segregation, and will decrease because of loss cone mechanism. The lower panel shows the relation between the average rate $\overline{\Gamma}$, the maximum rate $\Gamma_{max}$ and MBH masses. For the larger MBH mass ($M\gtrsim10^7M_\odot$), the rates have a sharp decrease because these systems don't evolve to a stable state in the cosmic time.}
    \label{fig:burst rate}
\end{figure}

Fig.~\ref{fig:burst rate} shows our numerically calculated event rate $\Gamma$. 
The upper panel shows the time evolution $\Gamma$ for various MBM mass $M$ and the lower panel shows the average $\overline{\Gamma}$ and maximum $\Gamma_{\rm max}$ event rate as functions of $M$.
The distribution of sBHs concentrate on deeper orbits due to the mass segregation effect and $\Gamma$ keeps growing to stable state with its maximum value $\Gamma_{\rm max}>1/$yr as seen in the upper panel.
After $\Gamma$ reaches its maximum $\Gamma_{\rm max}$, a slow-decaying plateau is observed as the evolution of sBHs is now dominated by the removal through the loss cone.
As $M$ increases, the average and maximum event rate both increase as seen in the lower panel.
For MBHs with mass $M\gtrsim 10^7 M_\odot$, the calculated event rate does not reach the plateau in cosmic time.
Therefore, the average and maximum event measured during our $13$~Gyr numerical evolution observe a sharp drop for $M\gtrsim 10^7 M_\odot$ in the lower panel. 

The expected number of the close encounters, summing up the contribution of all MBHs within redshift $z$ can be estimated as
\begin{equation}
N_z  = \zeta_{\rm WD}\int^z_0 dz \int  d \log M \frac{1}{1+z}\frac{d V_z}{d z}\xi(M) \mu(M) \Gamma \label{eq:Nz}
\end{equation}
where $\xi=d n/d\log M$ is the mass function of MBHs with $n$ being the number density of MBHs in the comoving frame and $\zeta_{\rm WD}$ is the average number of orbiting WDs per MBH. QPE observations imply the existence of one or more stellar-mass objects within a few hundred gravitational radii for at least $10\%$ of MBHs \citep{Chakraborty:2025}, as not all ${\rm EMRI}+{\rm disk}$ systems produce observable QPEs. Assuming that WDs make up a fraction $\sim 10\%$ of the stellar population as in the local universe \citep{Gaia:2021}, we conservatively estimate $\zeta_{\rm WD}\sim 0.01$.
We adopt a phenomenological mass function for the MBH
\begin{align}
\xi(M) = 0.005 \left(\frac{M}{3\times 10^6\, M_\odot}\right)^{\,\beta} {\,\rm Mpc}^{-3}\,.
\end{align}
We consider two choices of the exponent $\beta$: $\beta \approx 0$ inferred from the observational rate of TDEs \citep{Yao:2023rbr,Lyu:2024gnk} and $\beta=-0.3$ from self-consistent MBH population model by \citet{Babak2017}.

\begin{figure}[h]
    \centering
    \includegraphics[width=1.0\linewidth]{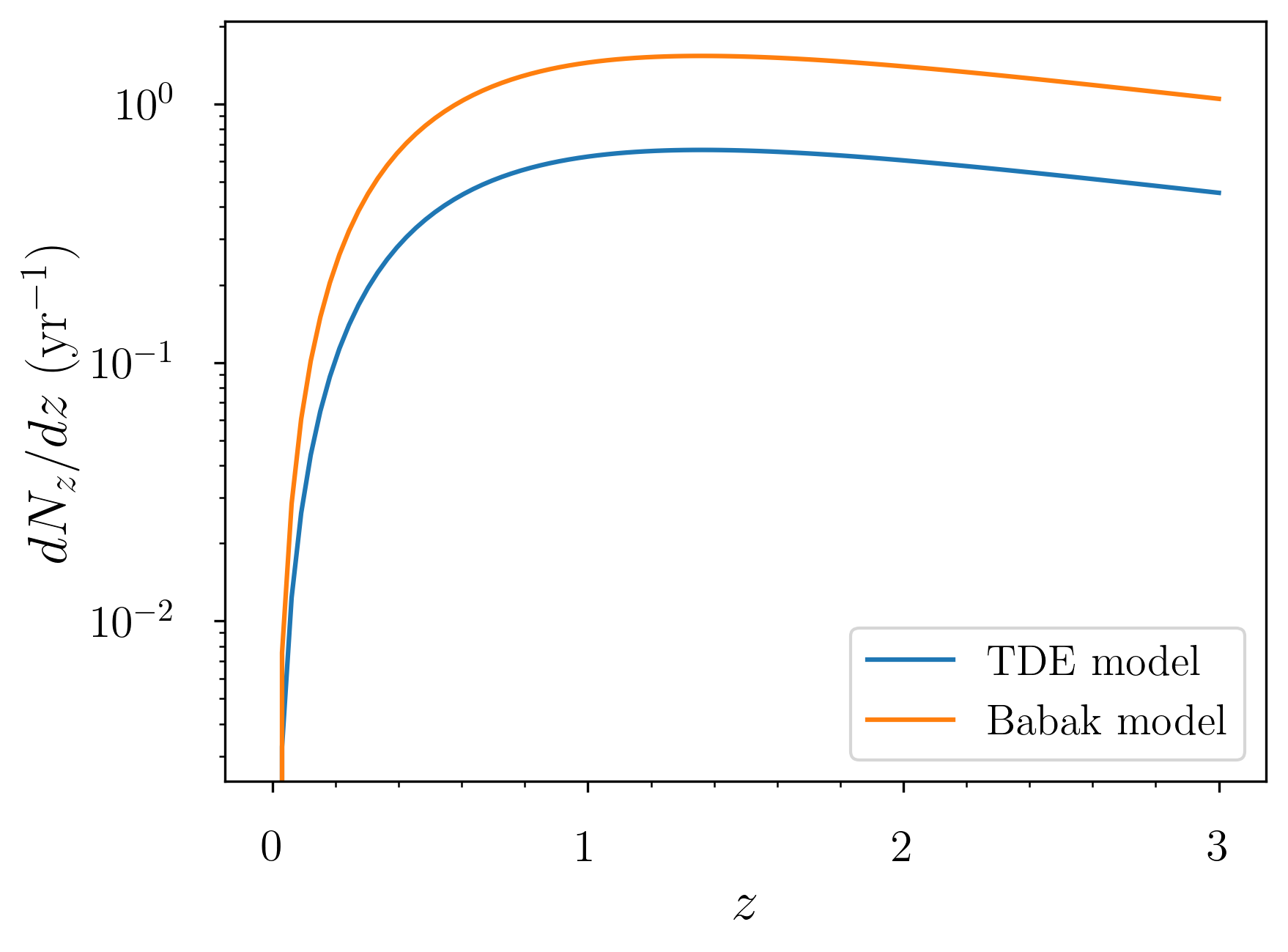}
    \label{dNz_dz}
    \hfill
    \includegraphics[width=1.0\linewidth]{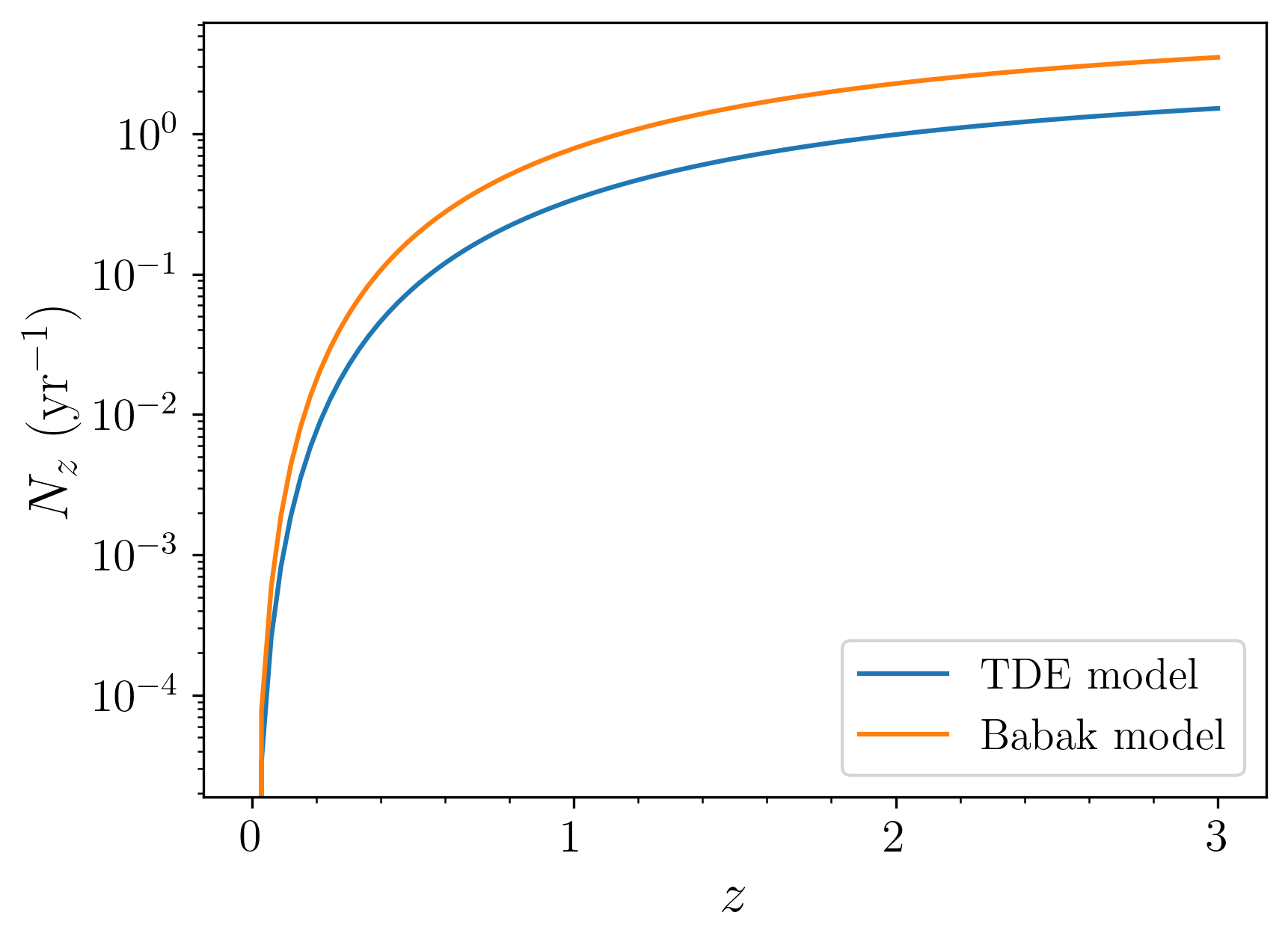}
    \caption{The upper panel shows the differential number of event rate with respect to redshift for two different MBH mass functions. It reaches the maximum value at $z\sim 1$. The lower panel shows the expected event rate within redshift for the two MBH mass functions. The rate reaches $\sim 0.1 / {\rm yr}$ within $z\leq1$ and $\sim 1 / {\rm yr}$ within $z\leq3$.}
    \label{fig:Nz}
\end{figure}

Fig.~\ref{fig:Nz} shows the total ($N_z$) and differential ($dN_z/dz$) number of encounter event per year as functions of redshift $z$ with $\zeta_{WD}=0.01$.
The blue lines are results using the mass function with $\beta=0$ (TDE model) and the orange lines with $\beta=-0.3$ (Babak model).
The event occurrence has a maximum at $z\sim1$.
We find for both mass functions, the total rate reaches $\ge 10 \zeta_{\rm WD} {\rm yr}^{-1}\sim 0.1$/yr within $z \le 1$ and $\ge 100 \zeta_{\rm WD}{\rm yr}^{-1} \sim 1$/yr within $z\le3$.

\section{Disruption Criterion}\label{sec:theory}
We model the encounter of the WD with the sBH as follows: a WD approaches the sBH from infinity with constant velocity $v_0$ and impact parameter $b$.
The velocity perturbation of the WD by the sBH after the encounter can be estimated as a Newtonian gravitational scattering process \citep{Binney:2008}
\begin{equation}
    \delta v = \frac{2Gm_{\rm sBH}}{bv_0}.
\end{equation}
Taking into account the finite size of the WD, the tidal force of the sBH after the encounter will create a velocity difference between the center of mass and material on the surface
\begin{equation}
    \Delta v \sim \frac{\partial (\delta v)}{\partial b}R_{\mathrm{WD}}.
\end{equation}
When the velocity difference exceeds $v_{\mathrm{esc}}\equiv\sqrt{2Gm_{\rm WD}/R_{\rm WD}}$ the escape velocity of the WD, the WD cannot remain gravitationally self-bound and is therefore disrupted.
Therefore, we can write the criterion of tidal disruption during the encounter as
\begin{equation}
    \bigg|\Delta v\bigg |\gtrsim v_{\mathrm{esc}},
\end{equation}
which gives
\begin{equation}\label{eq:criterion}
    b\lesssim b_{\rm{cr}}\equiv R_{\rm WD}\left(\frac{v_{\mathrm{esc}}}{v_0}\frac{m_{\rm sBH}}{m_{\rm WD}}  \right)^{1/2}.
\end{equation}

The magnitude of $b_{\mathrm{cr}}$ is on the order of $R_{\mathrm{WD}}$, which is much larger than the Schwarzschild radius of the sBH, justifying our estimation using Newtonian gravity.

After being tidally disrupted, the fate of the WD material falls into three categories. Part of the material will be captured by the sBH and eventually accreted onto the sBH. We denote this mass as $m_{acc}$. Part of the material may remain gravitationally self-bound, and we denote the mass of the self-gravitating remnants as $m_b$. The rest of the material will become unbound of the WD$+$sBH system, we denote the mass of the unbound debris as $m_{ub}$, 
\begin{equation}\label{eq:m_ub}
m_{ub}=m_{\rm WD}-m_{acc}-m_b.
\end{equation}

\section{Numerical Simulations}\label{sec:hydro}
\subsection{Setup}
To study how the different mass components depend on $b$ and $v_0$, we perform a series of numerical simulations to study the close encounter event.
We modify the ENZO \citep{enzo:2019JOSS} code to solve the Newtonian hydrodynamical equations in 3D with gravitational force from the sBH and self gravity from the WD material. No magnetic field is included in our simulations.

\begin{figure}[h]
    \centering
    \includegraphics[width=\linewidth]{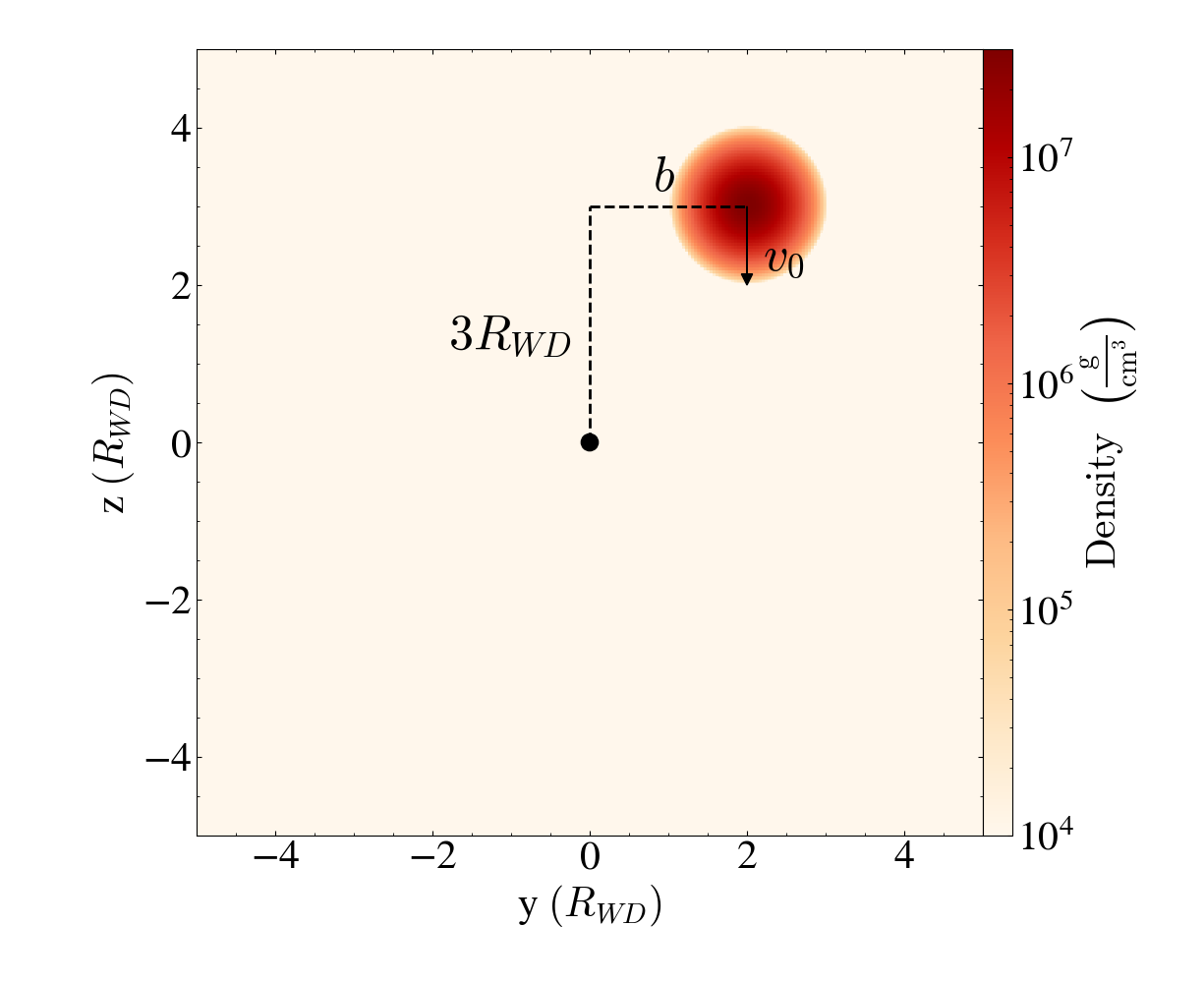}
    \caption{Initial condition for the simulations. The sBH of $10M_\odot$ is placed at the center shown as the black dot. The WD of $1M_\odot$ starts with impact parameter $b$ along the $y$-direction and moves with $v_0$ along the $z$-direction.}
    \label{fig:initial}
\end{figure}

The initial conditions for our numerical simulations are shown in Fig.~\ref{fig:initial}.
A $10M_\odot$ sBH is placed in the center of the computational box. 
A non-spinning WD of $1M_\odot$ with radius $R_{\rm{WD}}=5557$~km moves in a straight line along the $-z$ direction with initial speed $v_0$ and impact parameter $b$ along the $y$ direction.
Initially, the distance between the WD and the BH along $z$ direction is set to $3R_{\text{WD}}$.

The WD in our simulations is modeled using a simple equation of state for a zero-temperature degenerate electron gas given by
\citet{ShapiroTeukolsky:1986bhwd}
\begin{eqnarray}
P_e &=& \frac{2}{3h^3}\int\limits_0^{p_F}\frac{p^2c^2}{\sqrt{p^2c^2+m_e^2 c^4}}4\pi p^2\mathrm{d}p.
\end{eqnarray}
Here $P_e$ is the electron degeneracy pressure, $p$ is the electron momentum and $p_F$ is the Fermi momentum, $m_e$ is the electron mass, $h$ and $c$ are the Planck constant and the speed of light.
Details of the WD structure, including temperature and chemical composition, are not the main focus of this paper and are neglected in our simulations.
The sBH is treated as a point-source sink particle which is kept fixed at the center of the computational box and gas entering the sBH's neighboring cells is removed and considered accreted.

Our computational box is set to be $10$ times the WD radius $R_{\mathrm{WD}}$ along each side resolved by $256$ grid cells.
Therefore, the radius of the WD is well resolved by $\sim 25$ cells.
The physical length of the cell length is much larger than the Schwarzschild radius of the sBH, so general relativity can be safely neglected.

The simulations stop when the WD material starts to leave the computational box, at $t_{\rm{end}}=7R_{\mathrm{WD}}/v_0$, of the order of a few seconds.
We ensure the measured $m_{acc}$ and $m_{b}$ converge to stationary values in a few snapshots at the end of the simulations.

Convergence tests are performed with twice the resolution to ensure consistent results.
We also double the box size with the same resolution and start with the WD $6R_{\rm WD}$ away from the sBH along the $z$-direction, obtaining consistent outcomes.

We perform a series of simulations with different initial velocity
\begin{equation}
    v_0/c=0.05,0.1,0.15, 0.2\nonumber
\end{equation}
and impact parameter
\begin{equation}
b/R_{\text{WD}} = 0,0.25,0.5,0.75,1,1.5,2,2.5,3.\nonumber
\end{equation}

\subsection{Simulation results}
\begin{figure*}
\centering
\begin{minipage}{0.3\textwidth}
\centering
\includegraphics[width=\linewidth]{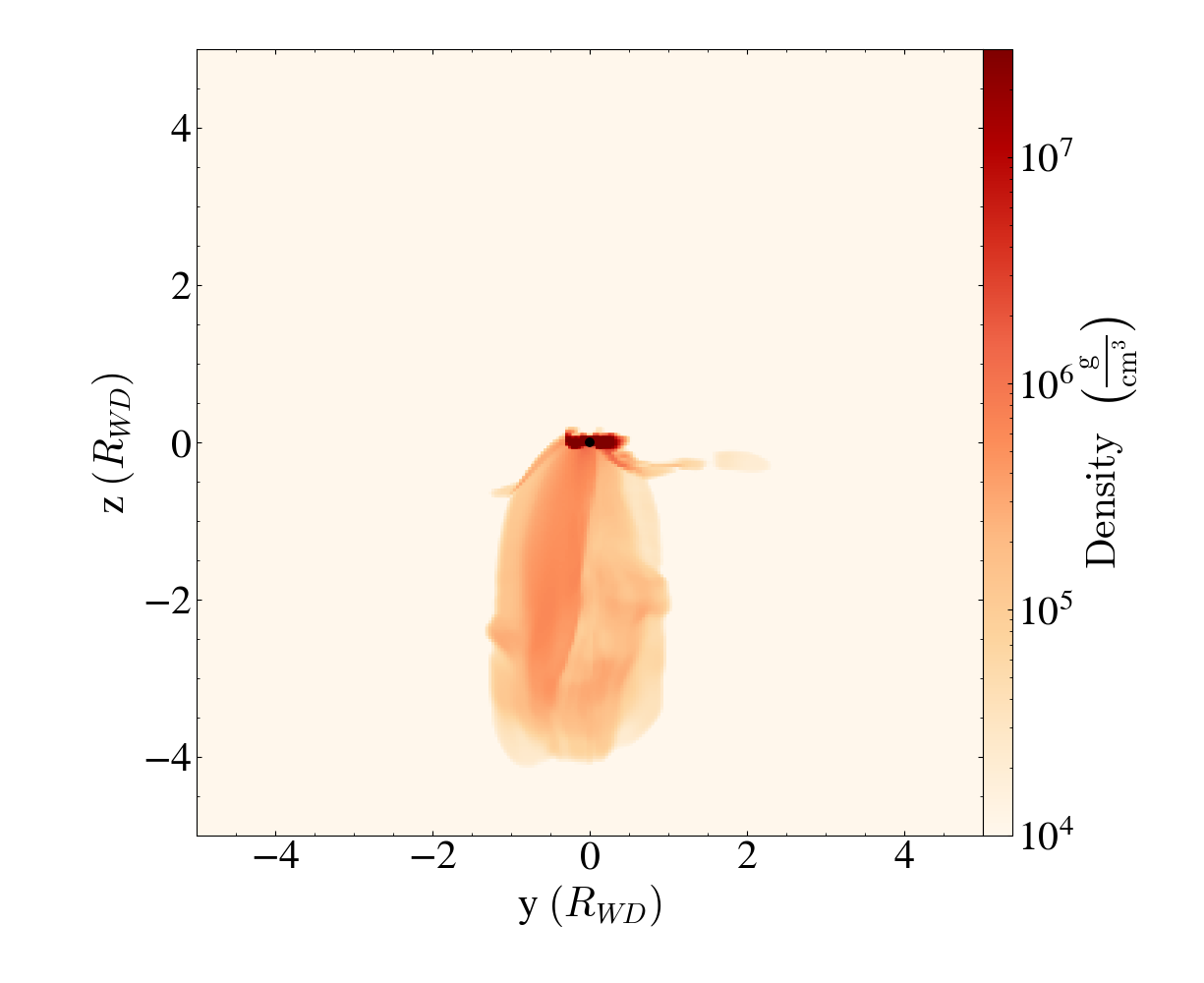} 
\textbf{(a)} $b = 0$
\label{fig:subim1_1}
\end{minipage}%
\begin{minipage}{0.3\textwidth}
\centering
\includegraphics[width=\linewidth]{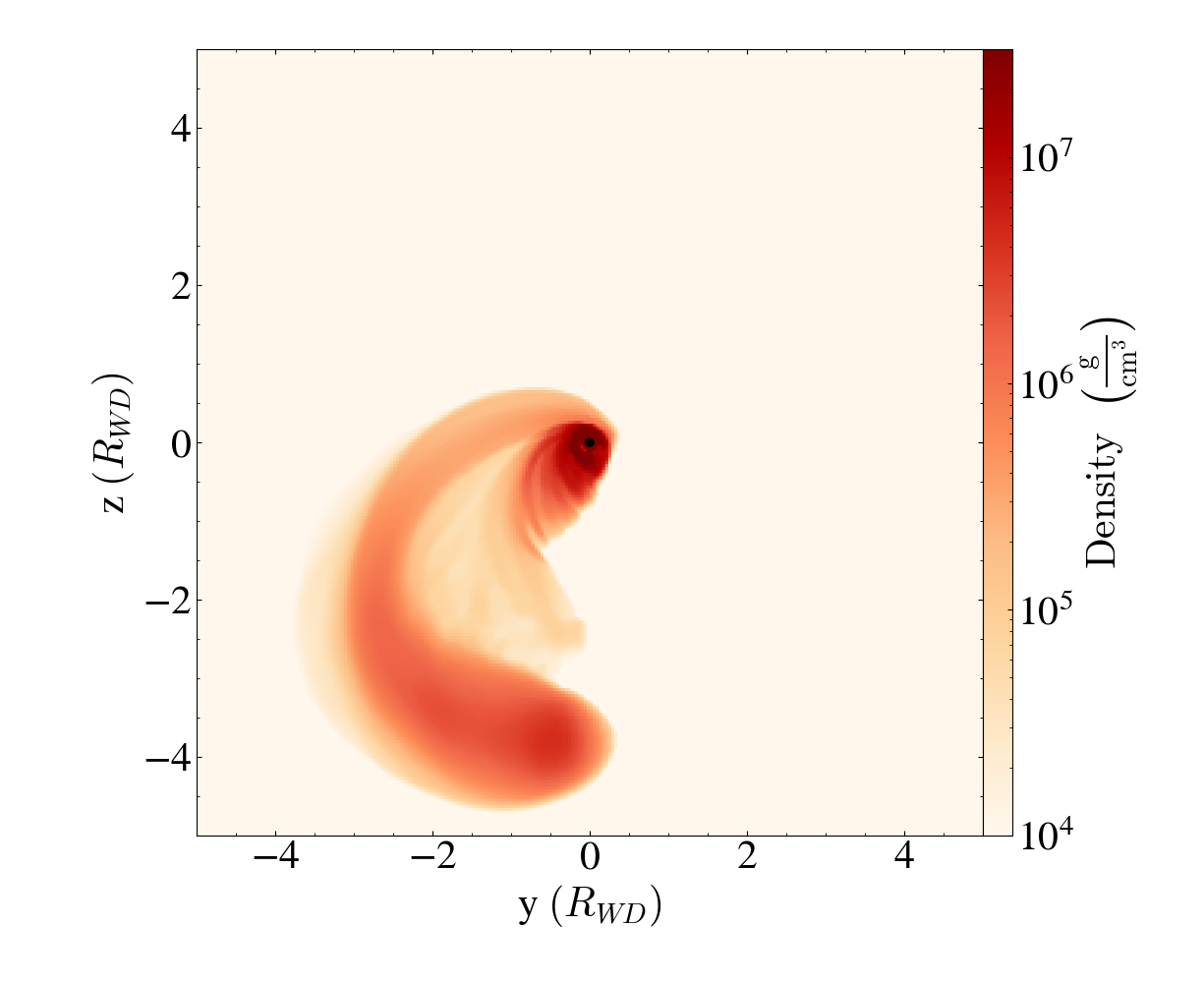}
\textbf{(b)} $b=0.5R_{\text{WD}}$
\label{fig:subim1_2}
\end{minipage}%
\begin{minipage}{0.3\textwidth}
\centering
\includegraphics[width=\linewidth]{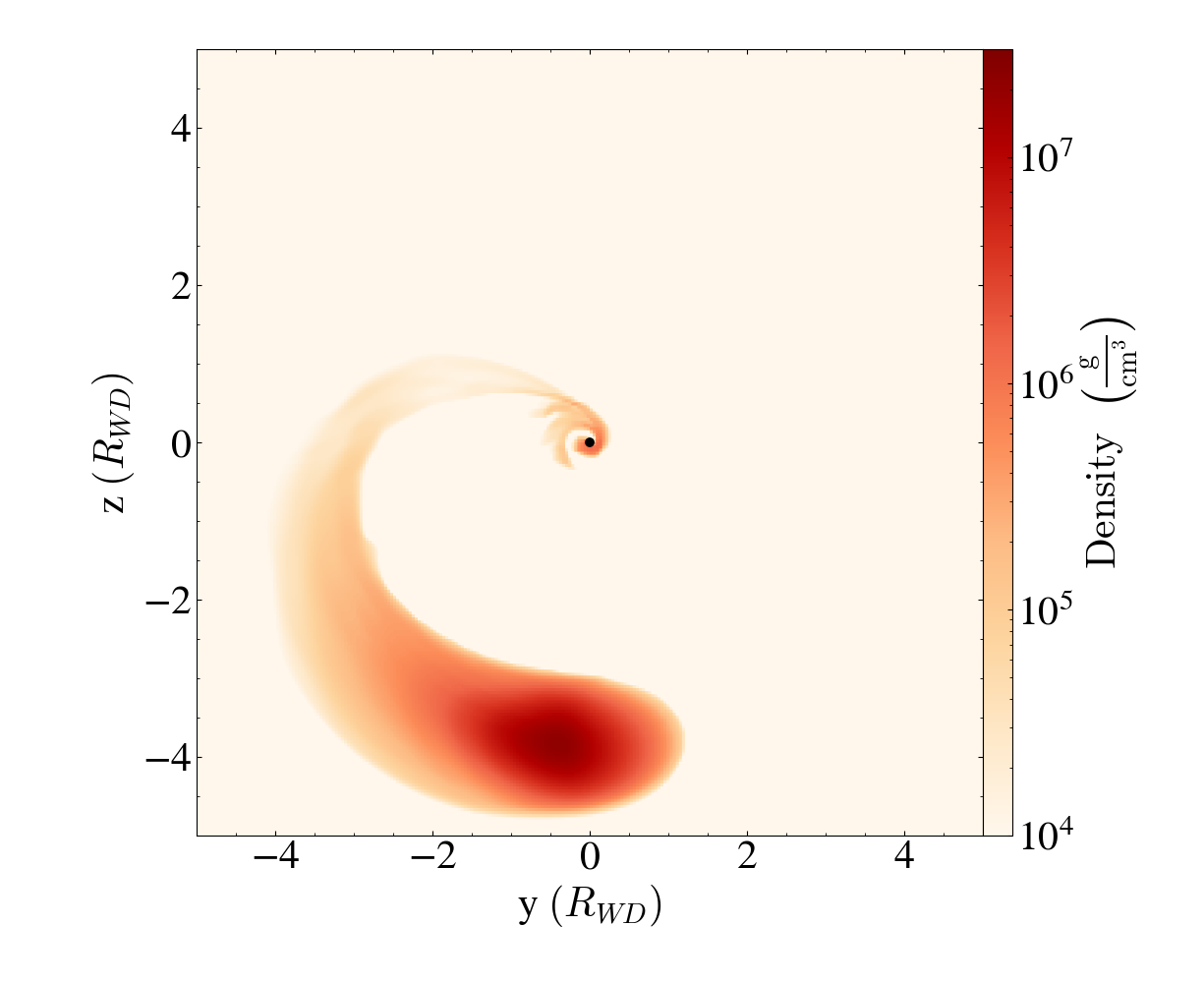}
\textbf{(c)} $b=R_{\text{WD}}$
\label{fig:subim1_3}
\end{minipage}

\vskip\baselineskip

\begin{minipage}{0.3\textwidth}
\centering
\includegraphics[width=\linewidth]{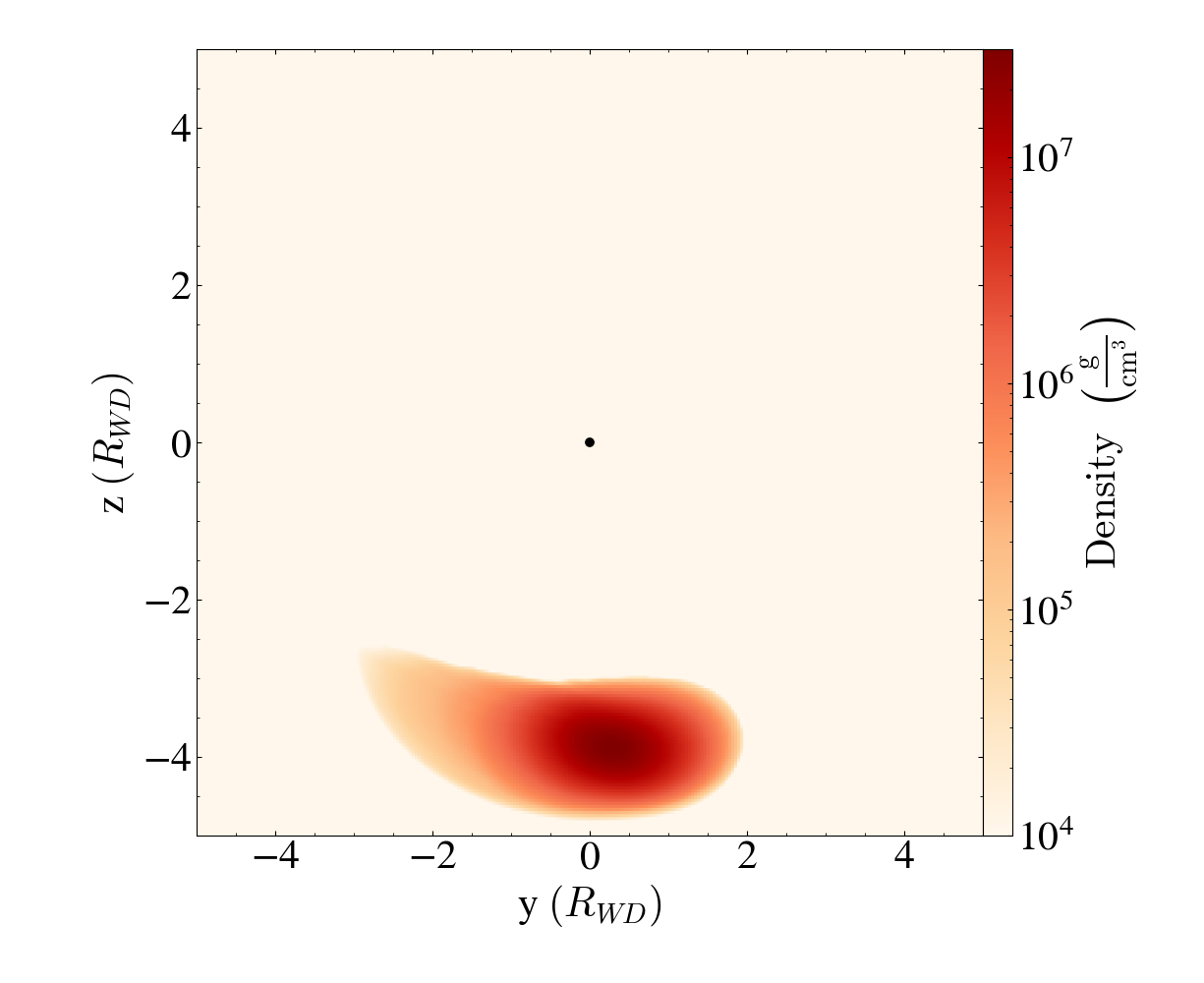}
\textbf{(d)} $b=1.5R_{\text{WD}}$
\label{fig:subim2_1}
\end{minipage}%
\begin{minipage}{0.3\textwidth}
\centering
\includegraphics[width=\linewidth]{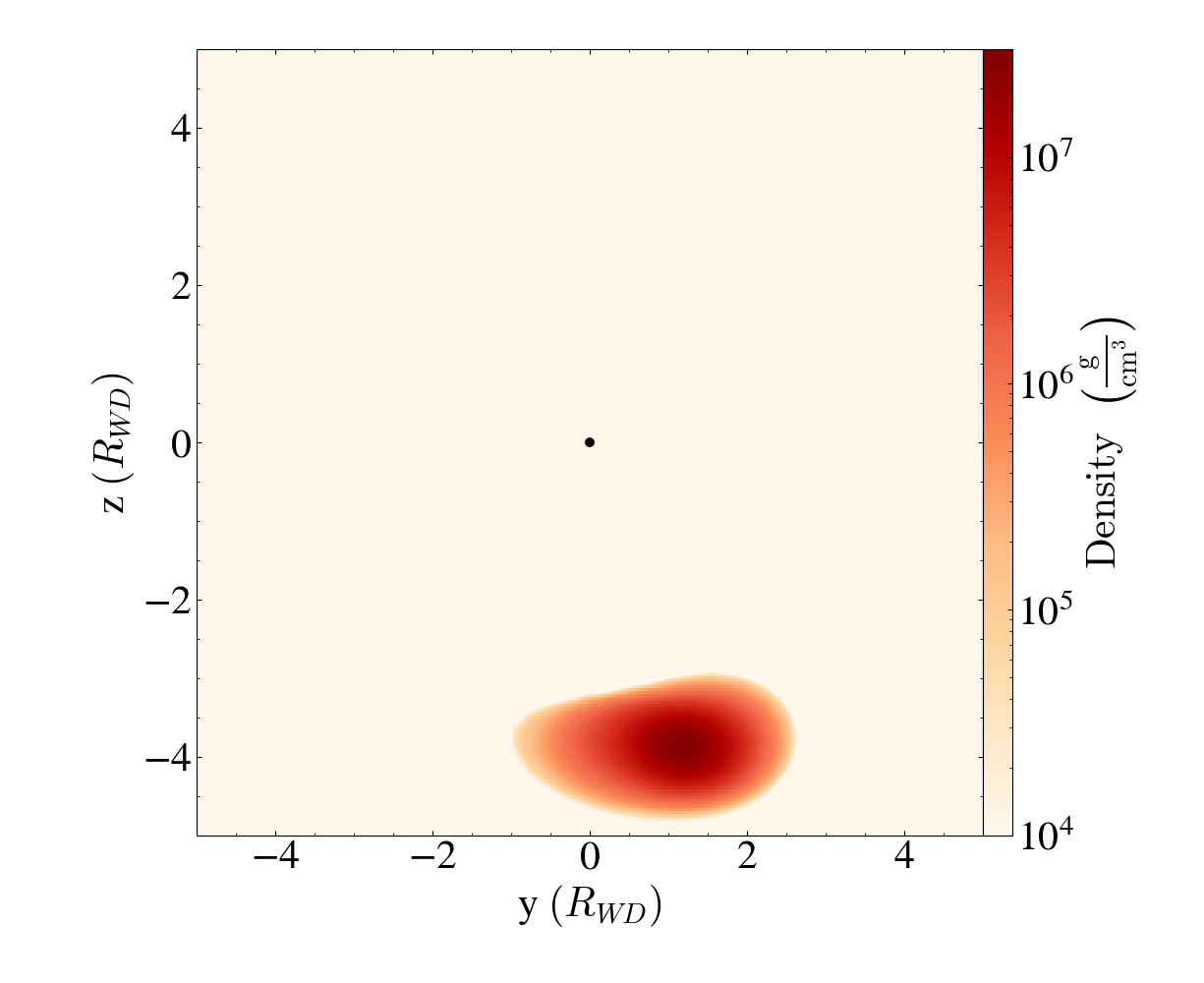}
\textbf{(e)} $b=2R_{\text{WD}}$
\label{fig:subim2_2}
\end{minipage}%
\begin{minipage}{0.3\textwidth}
\centering
\includegraphics[width=\linewidth]{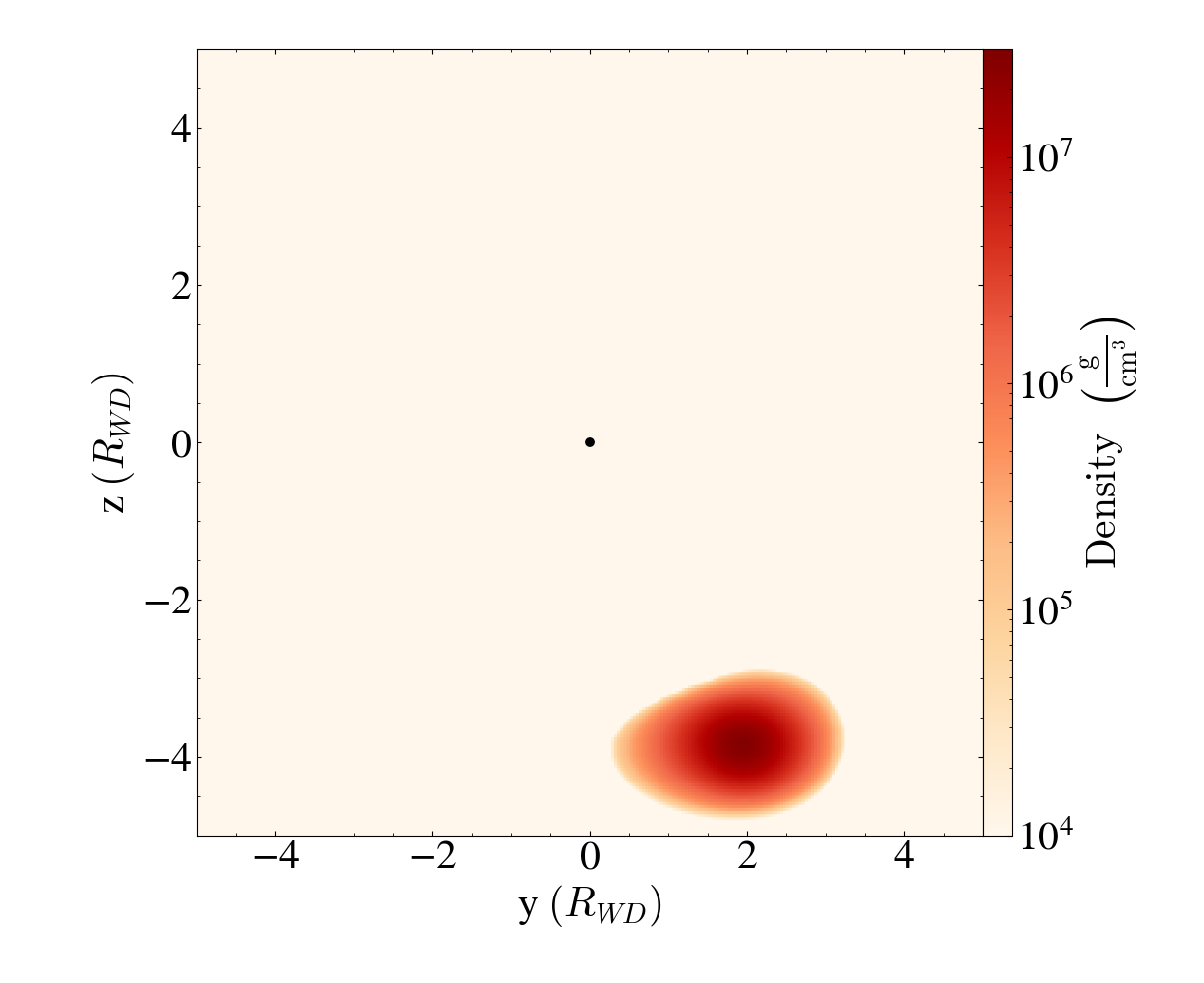}
\textbf{(f)} $b=2.5R_{\text{WD}}$
\label{fig:subim2_3}
\end{minipage}
\caption{Snapshots of gas density at $t=1.26$~s with $v_0=0.1c$ and different impact parameter $b$.}
\label{fig:snapshot}
\end{figure*}

\begin{figure*}
\centering
\begin{minipage}{0.3\textwidth}
\centering
\includegraphics[width=\linewidth]{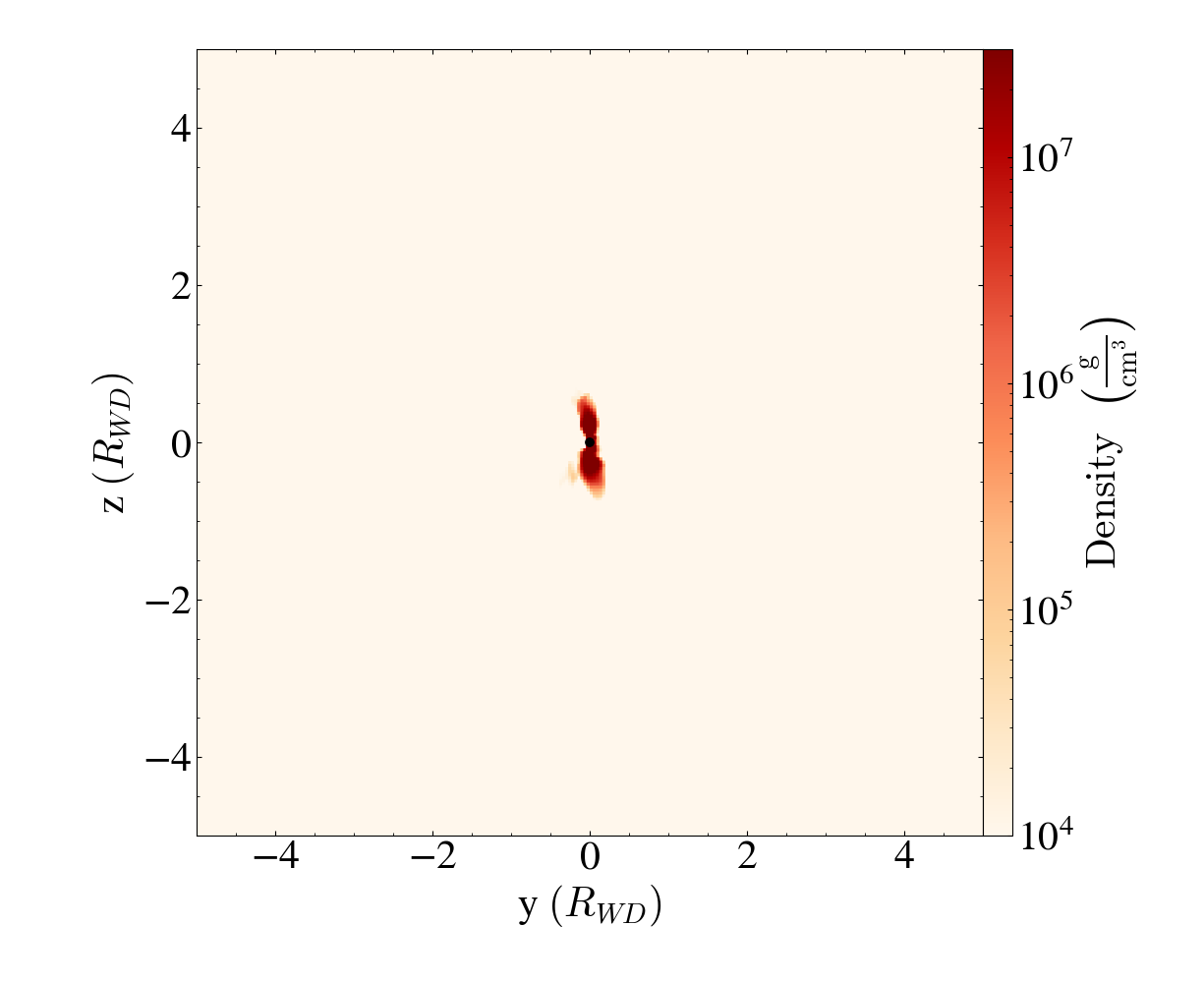} 
\textbf{(a)} $b=0$
\label{fig:subim3_1}
\end{minipage}
\begin{minipage}{0.3\textwidth}
\centering
\includegraphics[width=\linewidth]{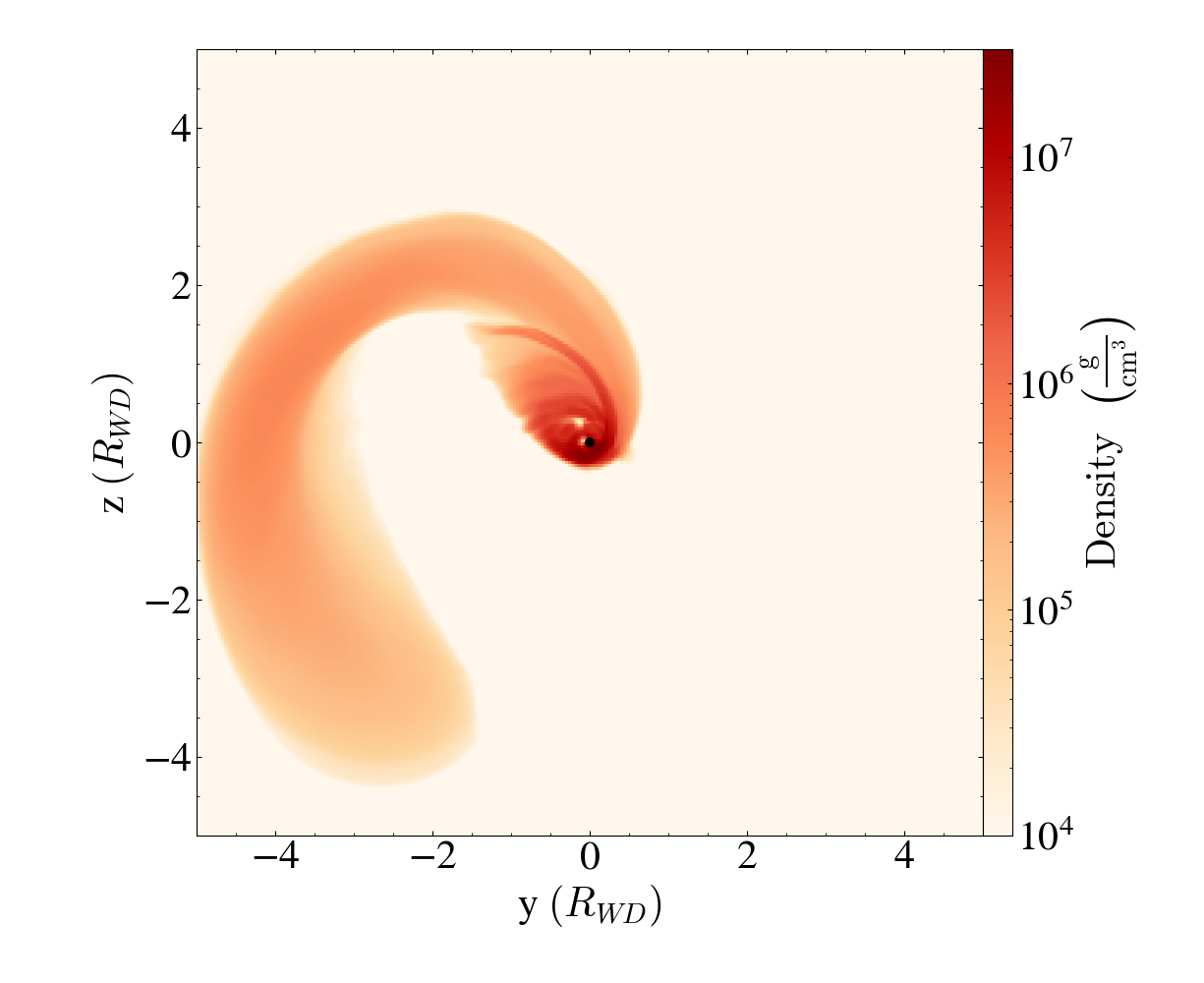}
\textbf{(b)} $b=R_{\text{WD}}$
\label{fig:subim3_2}
\end{minipage}
\begin{minipage}{0.3\textwidth}
\centering
\includegraphics[width=\linewidth]{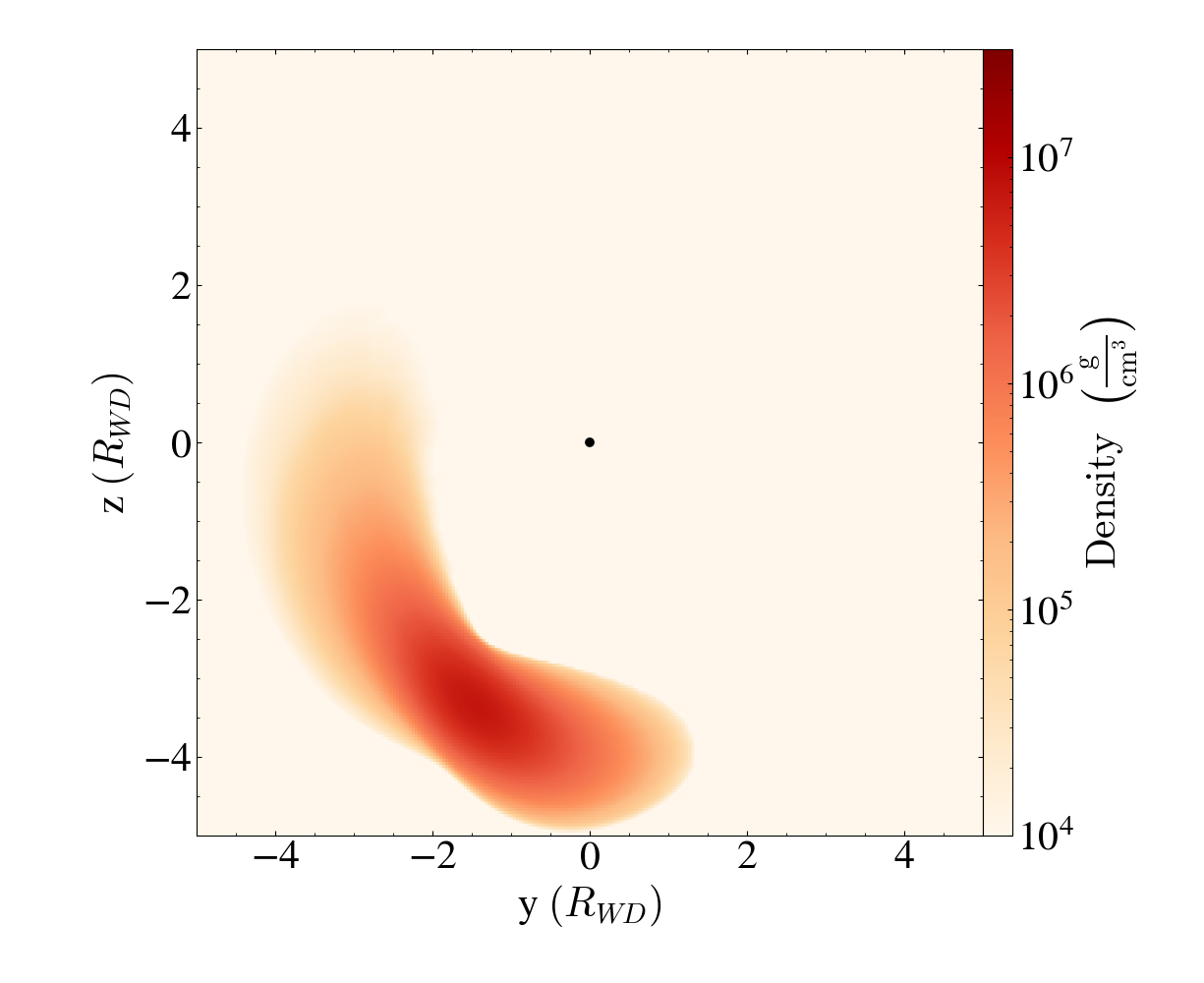}
\textbf{(c)} $b=2R_{\text{WD}}$
\label{fig:subim3_3}
\end{minipage}
\caption{Snapshots of gas density at $t=2.51$~s for $v_0=0.05c$ and different $b$.}
\label{fig:snapshot2}
\end{figure*}

\begin{figure*}
\centering
\begin{minipage}{0.3\textwidth}
\centering
\includegraphics[width=\linewidth]{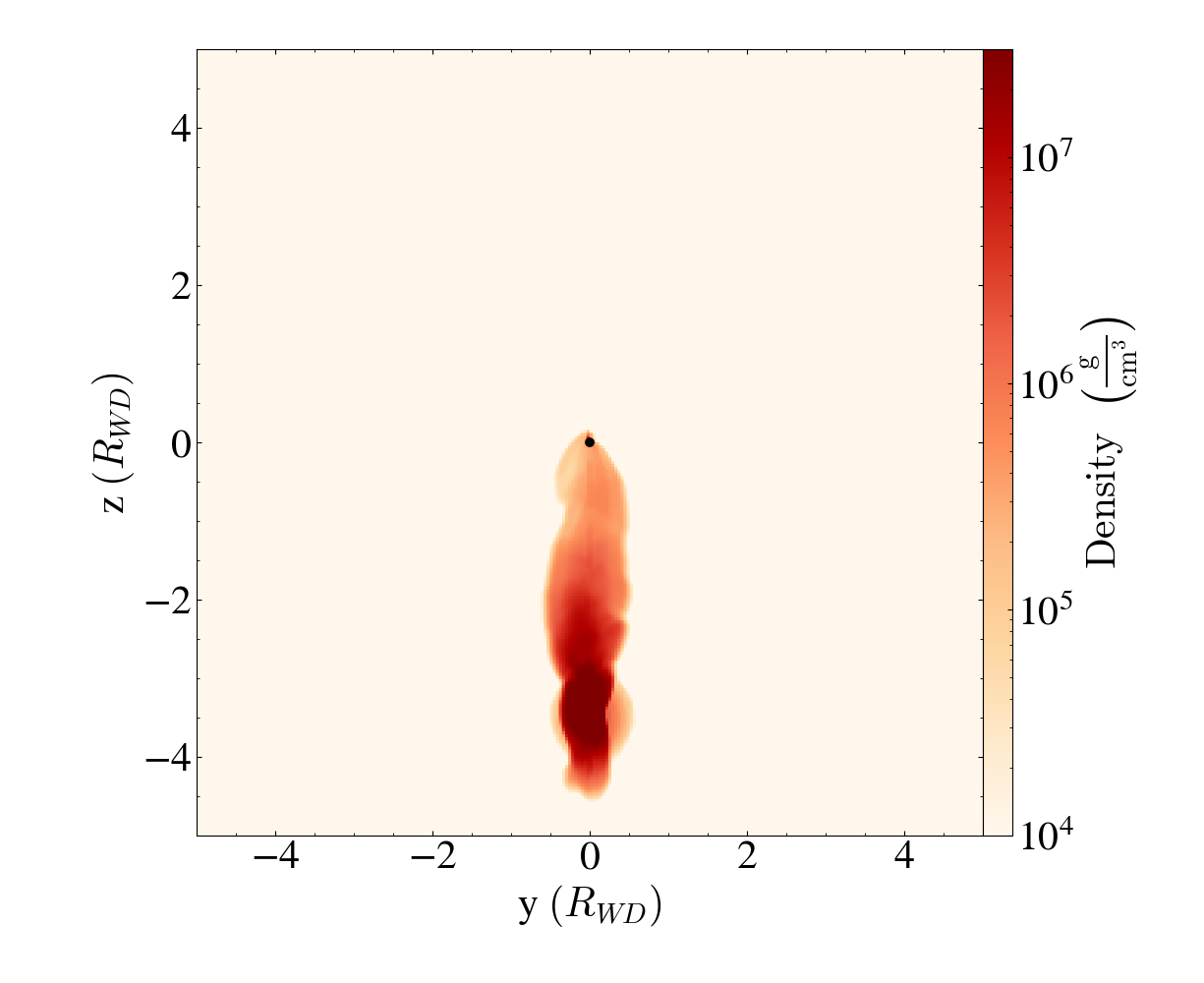} 
\textbf{(a)} $b=0$
\label{fig:subim1}
\end{minipage}
\begin{minipage}{0.3\textwidth}
\centering
\includegraphics[width=\linewidth]{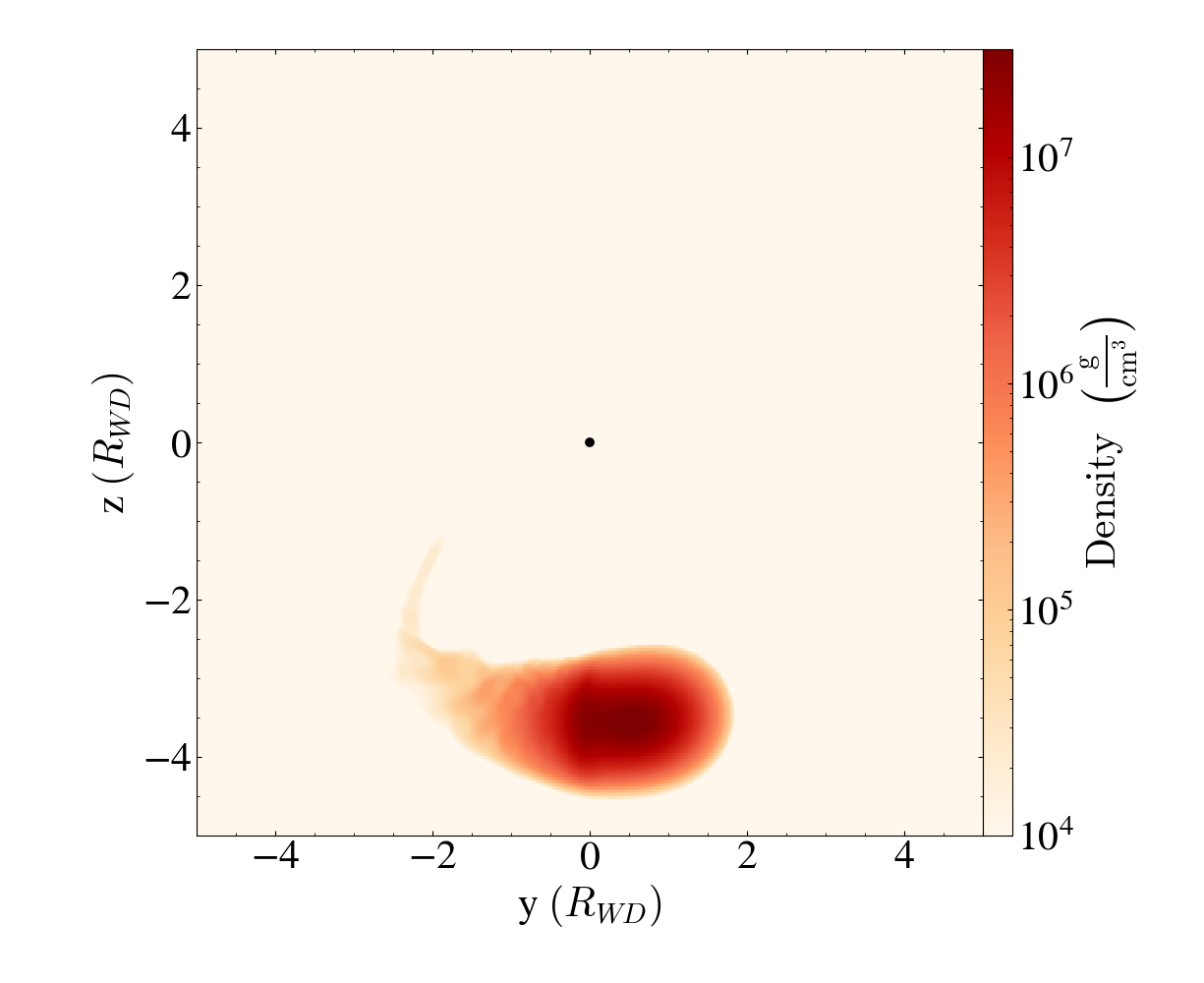}
\textbf{(b)} $b=R_{\text{WD}}$
\label{fig:subim2}
\end{minipage}
\begin{minipage}{0.3\textwidth}
\centering
\includegraphics[width=\linewidth]{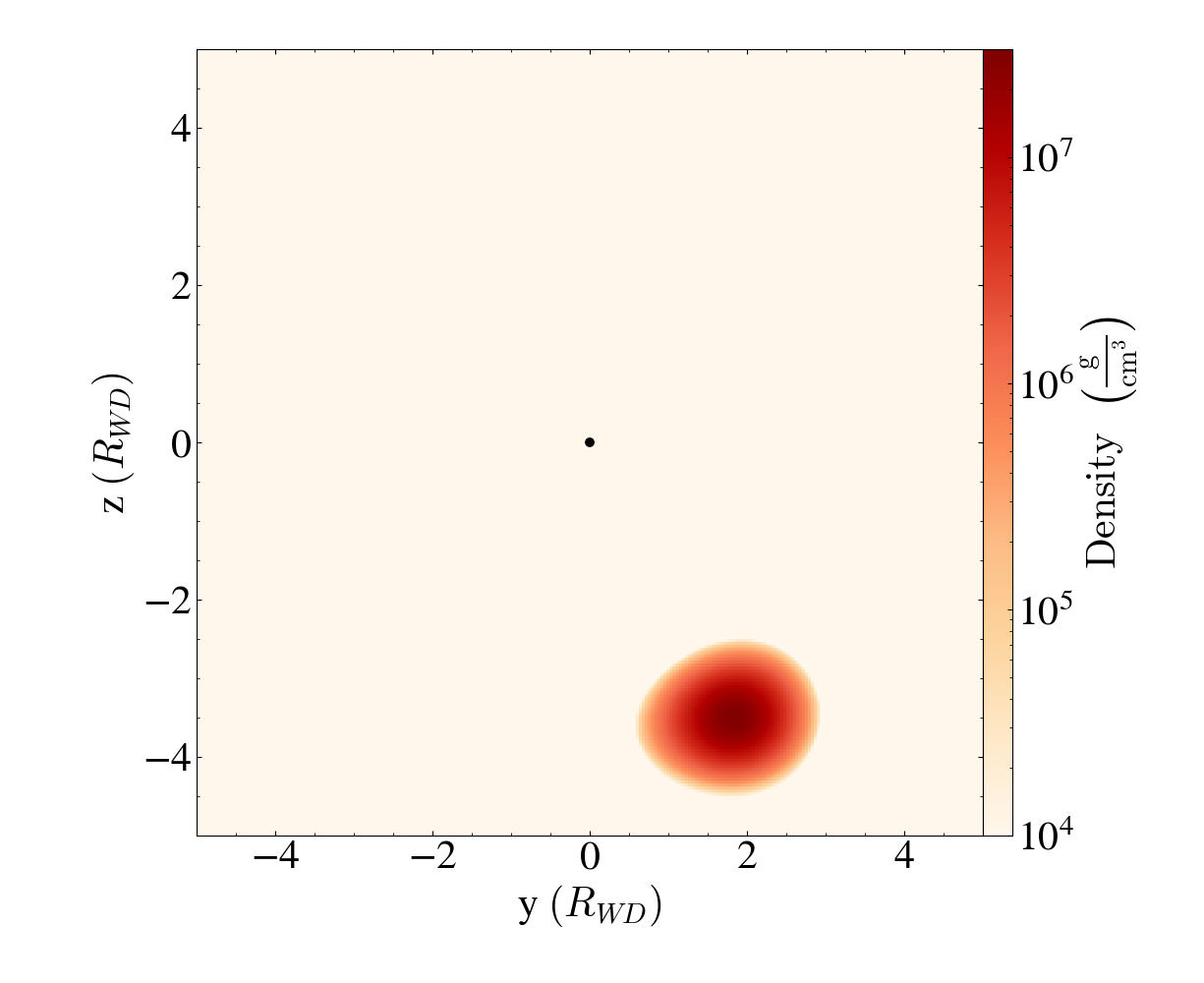}
\textbf{(c)} $b=2R_{\text{WD}}$
\label{fig:subim3}
\end{minipage}
\caption{Snapshots of gas density at $0.6$~s for $v_0=0.2c$ and different $b$.}
\label{fig:snapshot3}
\end{figure*}

Fig.~\ref{fig:snapshot} shows snapshots of the density distribution at $t=1$~s for $v_0=0.1c$ and various impact parameter $b$.
For $b>1.5R_{\rm WD}$ (the lower panels), the WD is slightly distorted with little mass falling onto the central sBH.
As $b$ decreases (the upper panels), the WD can get closer to the sBH, and begins to be tidally disrupted, and has part of its mass captured by the sBH, forming a disk at the center.
For $b=0$ the head-on collision case, the WD is totally disrupted as shown in the top left panel.
Figs.~\ref{fig:snapshot2} and \ref{fig:snapshot3} are snapshots for $v_0=0.05c$ at $1.6$~s and $v_0=0.2c$ at $0.5$~s which exhibit similar trends.

\begin{figure}[h]
    \centering
    \includegraphics[width=\linewidth]{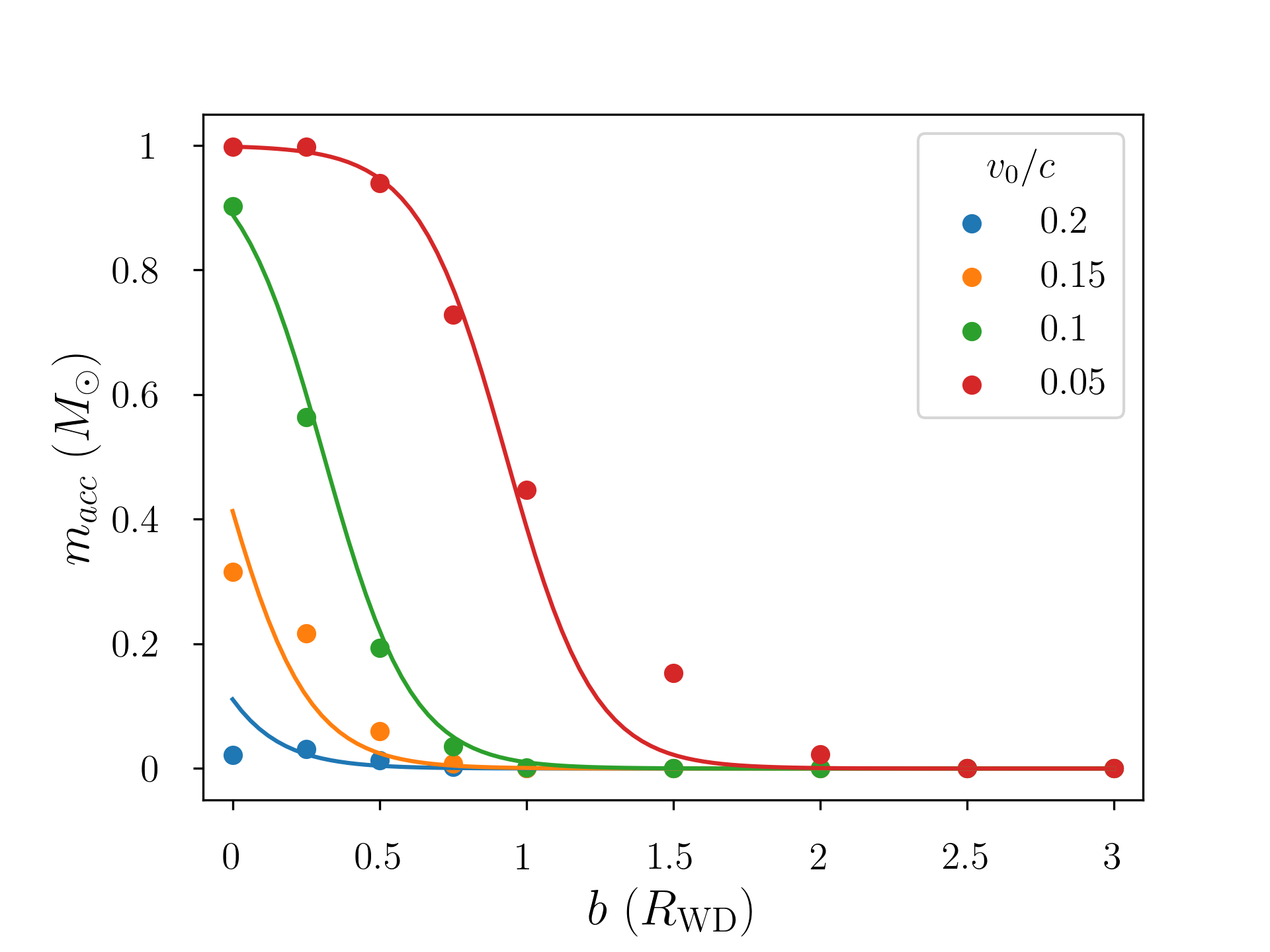}
    \caption{Relation between $m_{acc}$, mass captured by the sBH, and impact parameter $b$ for various initial velocity $v_0$. Dots are simulation results and lines are fitted lines from Eq.~\ref{eq:fit1}.
    }
    \label{fig:mass_accreted}
\end{figure}

To measure $m_{acc}$, the mass captured by the sBH, we employ the simple criterion that the total gas specific energy (sum of the kinetic energy and gravitational energy due to the central sBH) is negative
\begin{equation}
    U_{\rm tot} = \frac{1}{2} v^2-G\frac{m_{\rm sBH}}{r}<0.
\end{equation}
Fig.~\ref{fig:mass_accreted} shows $m_{acc}$ as a function of impact parameter $b$ for different $v_0$.
Dots are simulation data and the lines are best fit lines using the formula
\begin{equation}\label{eq:fit1}
    \frac{m_{acc}}{M_\odot} = \frac{1}{1+\alpha(v_0)\exp\left(3b/20R_{\rm WD}\right)},
\end{equation}
with 
\begin{equation}
    \alpha(v_0)=1.25\times 10^5\left(\frac{v_0}{c}\right)^6. \nonumber
\end{equation}
For fixed $v_0$, $m_{acc}$ increases with decreasing $b$, as a smaller $b$ allows the WD to approach the sBH more closely, resulting in more material being captured.
For the smallest $v_0=0.05c$ with $b<0.5 R_{\mathrm{RW}}$, almost all the WD material will be captured by the sBH.
As $v_0$ increases, the original WD material has higher kinetic energy and it becomes more difficult for the sBH to capture the material. 
Therefore $m_{acc}$ drops with increasing $v_0$.
For the largest velocity $v_0=0.2c$, although the head-on collision totally disrupts the WD as seen in Fig.~\ref{fig:snapshot3}, $m_{acc}$ remains small.

To measure $m_b$, the mass of the self-gravitating remnants, we first calculate the center-of-mass position and velocity $\vec{v}_{CM}$ for the material not captured by the sBH.
Then the peculiar velocity relative to the center of mass is given by $\vec{v}_p\equiv \vec{v}-\vec{v}_{CM}$.
$m_b$ is therefore the total mass with negative specific energy
\begin{equation}
    U_{b}=\frac{1}{2}v_p^2+\Phi<0
\end{equation}
where $\Phi$ is the gravitational potential of the material not captured by the sBH.

\begin{figure}[h]
    \centering
    \includegraphics[width=\linewidth]{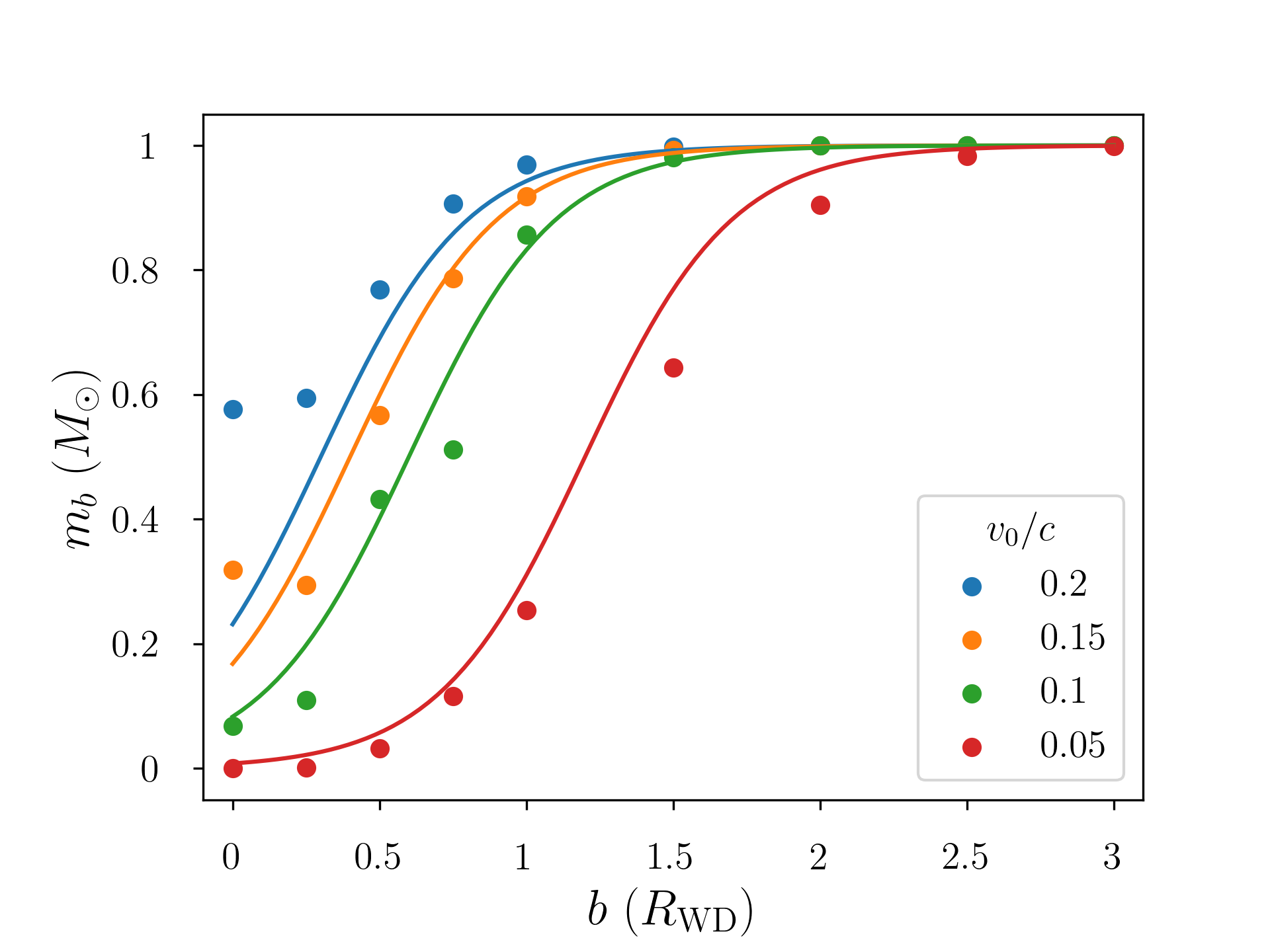}
    \caption{Relation between $m_{b}$, mass of the self-gravitating remnants, and impact parameter $b$ for various initial velocity $v_0$. Dots are simulation results and lines are fitted curves from Eq.~\ref{eq:fit2}.}
    \label{fig:mass_bound2}
\end{figure}

Fig.~\ref{fig:mass_bound2} shows $m_b$ as a function of $b$ for various $v_0$.
The solid lines are the best-fit curves following
\begin{equation}\label{eq:fit2}
    \frac{m_b}{M_\odot}=\frac{1}{1+\exp\left(\beta(v_0)-b/R_{WD}\right)}
\end{equation}
where
\begin{equation}
     \beta(v_0)=0.06\frac{c}{v_0}. \nonumber
\end{equation}
$m_b$ increases with $b$ and $v_0$, as the WD starts further away or with larger initial kinetic energy, the influence of the sBH decreases and more material will remain gravitationally self-bound after the encounter.

\begin{figure}[h]
    \centering
    \includegraphics[width=\linewidth]{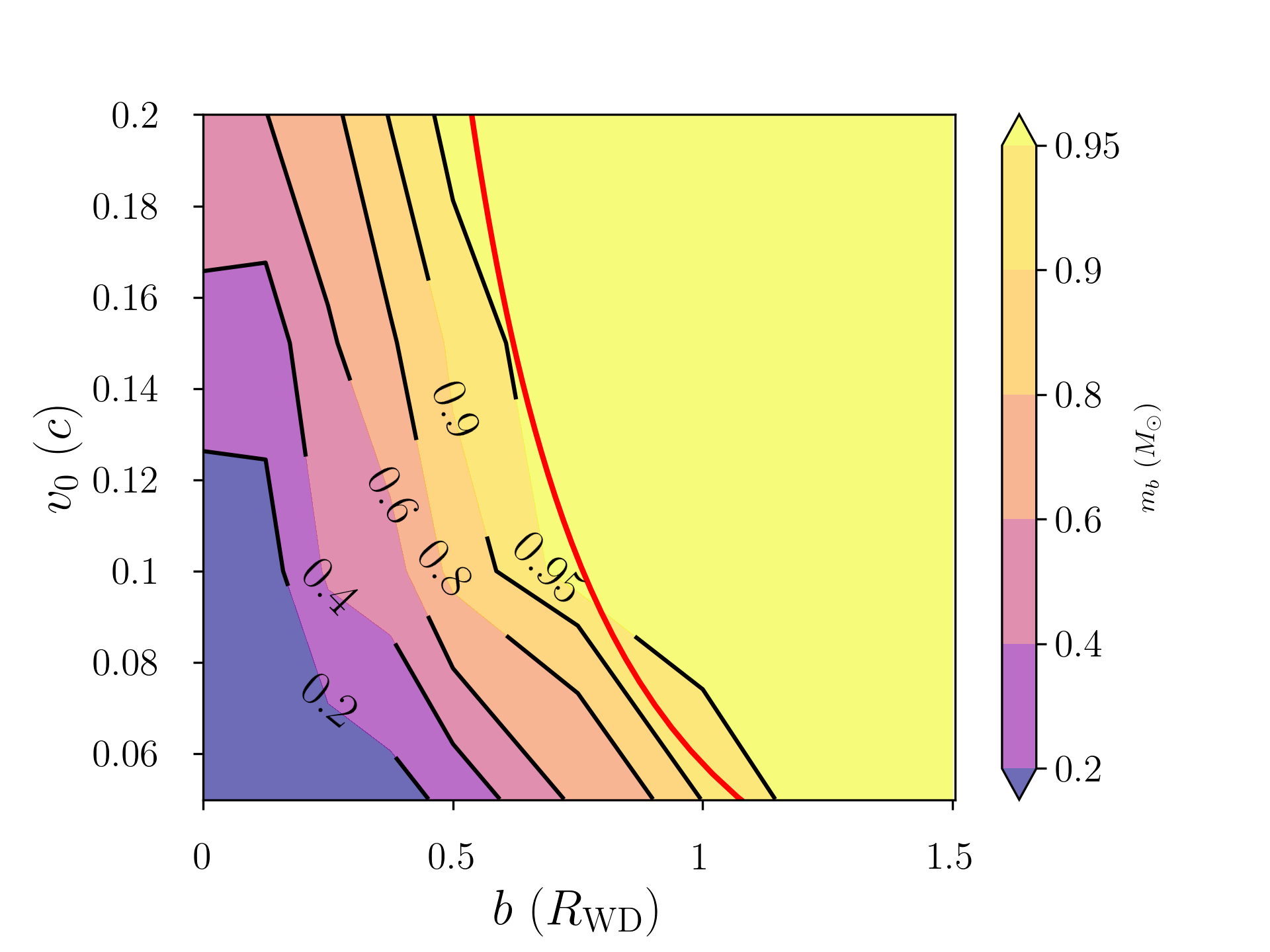}
    \caption{Contours of $m_{b}$ in the 2D parameter space spanned by $b$ and $v_0$.
    Black lines are contours of constant $m_b$ with the numbers indicating the value of $m_b$.
    Filled colors stand for regions of different $m_b$ value bins shown by the color map on the right.
    The red line draws the relation between $v_0$ and $b_{\mathrm{cr}}$ from Eq.~\eqref{eq:criterion}. 
    $b>b_{\mathrm{cr}}$ in regions right of the red line where the WD is not disrupted with $>95\%$ of the initial material remaining self-gravitating. $b<b_{\mathrm{cr}}$ in regions left of the red line where $m_b$ drops as the WD is significantly disrupted.}
    \label{fig:mass_bound}
\end{figure}

To test the validity of the simple estimation Eq.~\ref{eq:criterion}, the contours of $m_{b}$ in the 2D parameter space spanned by $b$ and $v_0$ are plotted in Fig.~\ref{fig:mass_bound}.
Black lines are contours of constant $m_b$ with the numbers indicating the value of $m_b$.
The contours are calculated from linear interpolation of the discrete simulation data.
The red line draws the relation between $v_0$ and $b_{\mathrm{cr}}$ from the theoretical estimation Eq.~\eqref{eq:criterion}. It generally follows the contour line of $m_b=0.95M_\odot$.
In regions right of the red line, $b>b_{\mathrm{cr}}$ is satisfied. There, the WD keeps more than $95\%$ of its mass and are only slightly perturbed by the sBH.
In regions left of the red line, $b<b_{\mathrm{cr}}$. $m_b$ starts to drop as $v_0$ and $b$ moving away from the red line towards the lower left corner indicating that the WD undergoes significant tidal disruption.
Eq.~\eqref{eq:criterion} provides a good criterion for whether the WD will be disrupted during the encounter.

\begin{figure}[h]
    \centering
    \includegraphics[width=\linewidth]{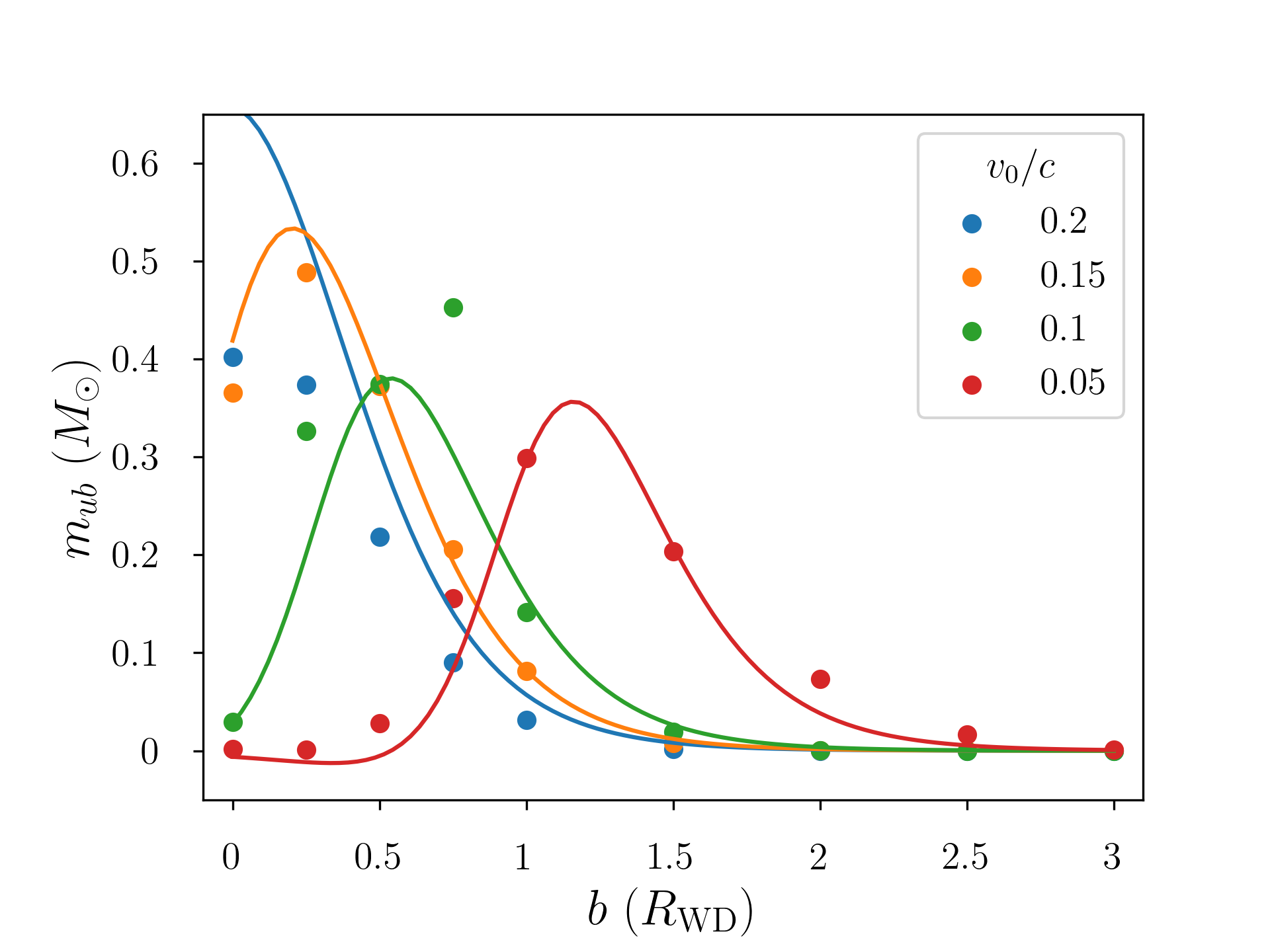}
    \caption{Relation between $m_{ub}$, the mass neither captured by the sBH nor gravitationally self-bound after the encounter, and impact parameter $b$ for various initial velocity $v_0$. Dots are simulation results and lines are calculated from Eq.~\ref{eq:m_ub} using fitted results from Eq.~\ref{eq:fit1} and \ref{eq:fit2}.}
    \label{fig:mass_unbound}
\end{figure}

Fig.~\ref{fig:mass_unbound} shows $m_{ub}$, the unbound mass, as a function of $b$ for various $v_0$.
The dots represent simulation data and the lines are fitted results from Eq.~\eqref{eq:fit1} and Eq.~\eqref{eq:fit2}.
For the small $v_0/c=0.05,0.1$ cases, $m_{ub}$ is small for both small $b$ (most material is captured by the sBH) and large $b$ (most material remains gravitationally self-bound).
Therefore, $m_{ub}$ shows a peak at $b\sim R_{\mathrm{WD}}$.
As $v_0$ increases, the original WD material have large kinetic energy.
Close encounters with very small $b$ or even head-on collision with $b=0$ may totally disrupt the WD, but the material will not be captured by the sBH. The material will stream freely in space. Therefore we observe $m_{ub}$ increases with $v_0$ at small $b$.

\begin{figure}[h]
    \centering
    \includegraphics[width=\linewidth]{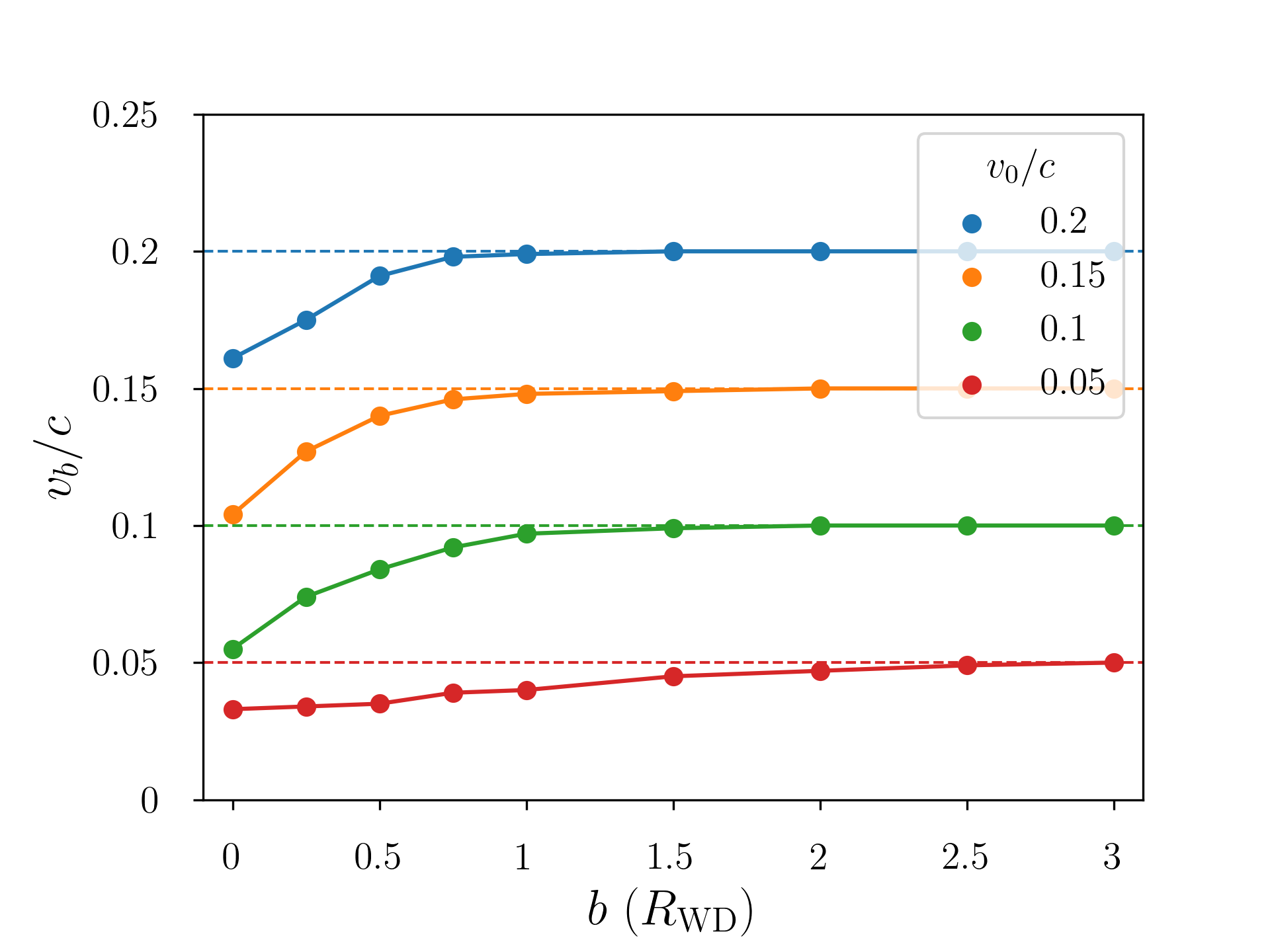}
    \caption{Relation between $v_b$, the center-of-mass velocity magnitude of the self-gravitating remnants after the passagge, and impact parameter $b$ for various initial velocity $v_0$. Dashed lines are $v_0$.
}
    \label{fig:vel_bound}
\end{figure}

Fig~\ref{fig:vel_bound} shows $v_b$, the center-of-mass speed of the self-gravitating remnants, as a function of $b$ for various $v_0$.
Dashed lines are $v_0$.
$v_b$ is not very different from $v_0$ except for small $b\lesssim R_{WD}$ when the WD can get very close to the sBH.
Therefore, the WD will remain on the same orbits around the MBH unless $b\lesssim R_{WD}$.

\begin{figure}[h]
    \centering
    \includegraphics[width=\linewidth]{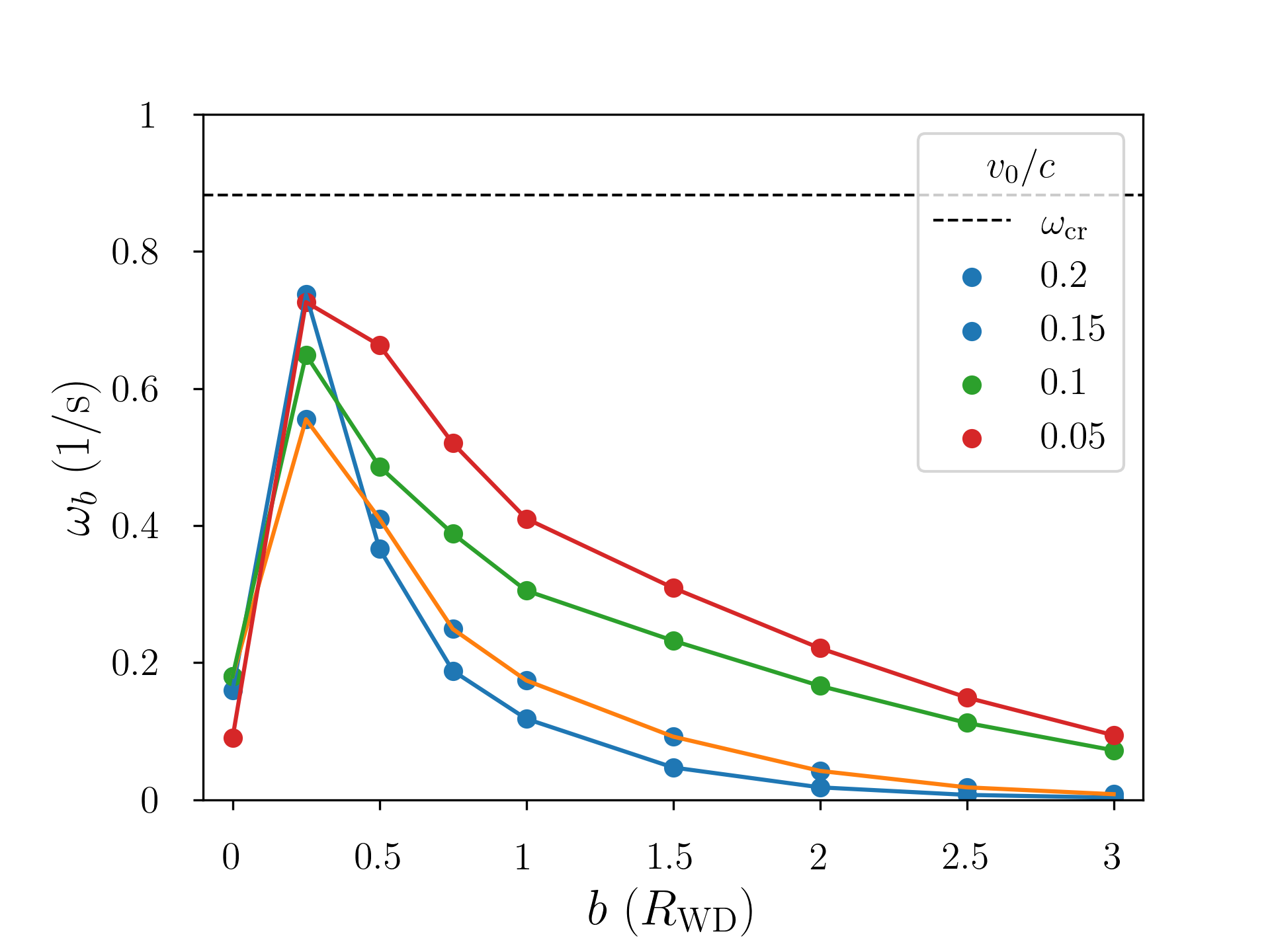}
    \caption{Relation between $\omega_b$, the spin rate of self-gravitating remnants, and impact parameter $b$ for various initial velocity $v_0$. The black dashed is the break-up angular velocity of the WD  $\omega_{cr}=\sqrt{Gm_{WD}/R_{WD}^3}$.}
    \label{fig:omega_bound}
\end{figure}

Fig~\ref{fig:omega_bound} shows $\omega_b$, the equatorial angular velocity of the self-gravitating remnants, as a function of $b$ for various $v_0$.
We calculate $\omega_b$ with respect to the axis along x-direction passing through the center of mass.
The black dashed line shows $\omega_{cr}\equiv\sqrt{Gm_{WD}/R_{WD}^3}$, the break-up angular velocity.
Due to the symmetry of our numerical setup, we do not expect the material to develop significant spin along other directions which is confirmed by the simulation data.
A spin rate $\omega_b<1$~s$^{-1}$ is developed after the encounter and decreases with increasing $b$ and $v_0$ as the influence of the sBH decreases.
The spin rate must drop for the head-on collision case $b=0$, as it is now mirror-symmetric in the x-direction.
Therefore, a peak in $\omega_b$ is expected at small, non-zero $b$.
The post encounter spin rates are all below the break-up angular velocity $\omega_{cr}$.

All the results from the simulation runs are listed in Table \ref{tab:summary}. 
Values below $10^{-6}$ are denoted $0$ in the table.

\begin{table*}[t]
\centering
\begin{tabular}{c|c|c|c|c|c|c}
  \hline
 $v_0/c$ & $b/R_{\rm{WD}}$ & $m_{acc}/M_\odot$ & $m_{b}/M_\odot$ &$m_{ub}/M_\odot$ & $v_b/c$ &$\omega_b$ (1/s)\\  
 \hline\hline
 \multirow{9}{*}{0.05} 
  & 0     & 0.998     & $10^{-5}$ & 0.002    & 0.033 & 0.089 \\
  & 0.25  & 0.997     & $10^{-3}$ & 0.003    & 0.034 & 0.726 \\
  & 0.5   & 0.94      & 0.03      & 0.03     & 0.035 & 0.663 \\
  & 0.75  & 0.73      & 0.12      & 0.16     & 0.039 & 0.522 \\
  & 1     & 0.45      & 0.25      & 0.30     & 0.004 & 0.413 \\
  & 1.5   & 0.15      & 0.64      & 0.20     & 0.045 & 0.309 \\
  & 2     & 0.04      & 0.90      & 0.07     & 0.047 & 0.221 \\
  & 2.5   & $2\times10^{-5}$ & 0.98      & 0.016    & 0.049 & 0.149 \\
  & 3     & 0         & 1         &$10^{-3}$ & 0.05 & 0.094 \\
 \hline
 \multirow{9}{*}{0.1} 
  & 0     & 0.90      & 0.07      & 0.03     & 0.055 & 0.182 \\
  & 0.25  & 0.56      & 0.11      & 0.32     & 0.074 & 0.649 \\
  & 0.5   & 0.19      & 0.43      & 0.37     & 0.084 & 0.486 \\
  & 0.75  & 0.04      & 0.51      & 0.45     & 0.092 & 0.388 \\
  & 1     & $10^{-3}$ & 0.86      & 0.14     & 0.097 & 0.305 \\
  & 1.5   & 0         & 1         & 0.02     & 0.099 & 0.232 \\
  & 2     & 0         & 1         & $5\times10^{-4}$     & 0.1 & 0.166 \\
  & 2.5   & 0         & 1         & 0        & 0.1   & 0.112 \\
  & 3     & 0         & 1         & 0        & 0.1   & 0.072 \\
   \hline
 \multirow{9}{*}{0.15} 
  & 0     & 0.32      & 0.32      & 0.37     & 0.104 & 0.183 \\
  & 0.25  & 0.21      & 0.29      & 0.49     & 0.127 & 0.555 \\
  & 0.5   & 0.06      & 0.57      & 0.37     & 0.14 & 0.41 \\
  & 0.75  & $8\times10^{-3}$& 0.79   & 0.21     & 0.146 & 0.249 \\
  & 1     & $10^{-4}$      & 0.92    & 0.08     & 0.148 & 0.174 \\
  & 1.5   & 0         & 0.99      & $8\times10^{-3}$ & 0.149 & 0.092 \\
  & 2     & 0         & 1         & $4\times10^{-5}$ & 0.15 & 0.042 \\
  & 2.5   & 0         & 1         & 0        & 0.15 & 0.018 \\
  & 3     & 0         & 1         & 0        & 0.15 & 0.008 \\
   \hline
 \multirow{9}{*}{0.2} 
  & 0     & 0.02      & 0.58      & 0.4    & 0.161 & 0.091 \\
  & 0.25  & 0.03      & 0.60      & 0.37    & 0.175 & 0.737 \\
  & 0.5   & 0.01      & 0.77      & 0.22     & 0.191 & 0.366 \\
  & 0.75  & $3\times10^{-3}$ & 0.9   & 0.1     & 0.198 & 0.188 \\
  & 1     & $10^{-5}$ & 0.97      & 0.03     & 0.2 & 0.118 \\
  & 1.5   & 0         & 0.998     & $2\times10^{-3}$     & 0.2 & 0.047 \\
  & 2     & 0         & 1         & 0        & 0.2 & 0.018 \\
  & 2.5   & 0         & 1         & 0        & 0.2 & 0.007 \\
  & 3     & 0         & 1         & 0        & 0.2 & 0.003 \\
  \hline
\end{tabular}
\caption{Summary of results from all the simulation runs. Values below $10^{-6}$ are counted as $0$ in the table.}
\label{tab:summary}
\end{table*}

\section{Observational Signatures}\label{sec:em}

\subsection{X-ray burst}
The material captured by the sBH circularizes and forms a transient accretion disk as seen in simulations.
This material will be accreted onto the sBH and produce thermal X-rays similar to those widely observed in X-ray binaries \citep{Dunn:2010}.
The thermal radiation of the accretion disk is unbeamed and limited by the Eddington luminosity $L_{Edd}$, and the disk temperature is given by \citet{Ulmer:1999}
\begin{align}
T_{\rm eff} \sim 8 \times 10^6 \left({\frac{M_{\text{sBH}}}{10 M_{\odot}}}\right)^{-1/4} {\rm K}.
\end{align}

The initial accretion rate will be super- or even hyper-Eddington.
During the WD tidal disruption event around $10^5 M_\odot$ black holes, the timescale for accretion to exceed the Eddington limit was found to be months to a year \citep{MacLeod:2014}.
Relativistic jets are generally expected for super-Eddington accretion  (e.g. \citet{Strubbe:2009}).
The power of the jets will depend on BH spin and magnetic flux as in
the Blandford-Znajek mechanism \citep{Blandford:1977} or by the collimation of the radiation field \citep{Sadowski:2015}. 
In both cases, the jet power is expected to be a fraction of $\dot{M}c^2$
(e.g. \citet{Krolik:2012,Tchekhovskoy:2013,Sadowski:2015}). 
As $\dot{M}c^2\gg L_{Edd}$, the X-ray emission from the jet can dominate over the thermal emission for viewers on the jet axis \citep{MacLeod:2014}.
Study of the detailed accretion and jet-launching process by simulations is beyond the scope of this paper, as magnetic fields are required with resolution down to the Schwarzschild radius of the sBH $r_s\sim10$~km. We leave this to future studies.

The circularization timescale is still uncertain but should be comparable to the fallback time at distance $b$ or even faster \citep{Dai:2015}
\begin{align}
 t_{\rm cir}(b) & = 2 \pi \sqrt{\frac{b^3}{G M_{\text{sBH}}}} \nonumber \\
& = 2.5 {\rm s} \left ( \frac{m_{\rm sBH}}{10 M_\odot}\right )^{-1/2} \left ( \frac{b}{R_{\rm WD}}\right )^{3/2},
\end{align}
and the viscous timescale of the disk is given by \citet{Shen:2019}
 \begin{align}
t_{\rm vis} & \sim 6.5 t_{\rm cir}\left( \frac{0.1}{\alpha} \right)^{-1} \left(\frac{h}{0.5} \right)^{-2} \nonumber \\
& \sim 16 {\rm s} \left( \frac{m_{\rm sBH}}{10 M_\odot}\right)^{-1/2} \left( \frac{b}{R_{\rm WD}}\right)^{3/2} \left( \frac{0.1}{\alpha}\right)^{-1} \left(\frac{h}{0.5} \right)^{-2}, 
 \end{align}
where $h$ is the aspect ratio of the disk and  $\alpha$ is the viscosity parameter.
During the super-Eddington transient accretion, the disk may be better described by the slim-disk scenario with moderate $h$ \citep{Wen:2020}. 
The circularization timescale describes the rise-up timescale of the transient, and the viscous time better characterizes the duration of the transient.
The average accretion rate can be estimated as $\dot{M} \sim  m_{acc}/t_{\rm vis}$. 
Assuming an efficiency $\eta \sim 0.1$ for the conversion of gravitational binding energy to radiation \citep{Ulmer:1999,Kremer:2021ApJ}), the jet power is $L_{X} \sim \eta m_{acc}c^2/t_{\rm vis}$.
For $b = R_{\rm WD}$ and $v_0=0.05c$, we have $m_{acc}=0.45M_\odot$ from Table~\ref{tab:summary}.
The accretion rate is $\dot{M}\sim 0.03M_\odot/\rm s$, and the jet power is $L_{X} \sim 10^{51}$~erg/s,

The mass accretion rate calculated above should be viewed as an upper limit.
\citet{Blandford:1999} proposed the adiabatic inflow–outflow
(ADIOS) model, where only a small fraction of the mass supplied at large radii is actually accreted.
The accretion rate is reduced by $(10r_s/r_d)^s$, where $r_s$ is the Schwarzschild radius of the black hole and $r_d$ is the disc radius, with the exponent $s$ between $0$ and $1$.
In our case $r_d$ is comparable to the WD radius $R_{\rm {WD}}$.
For the maximal reduction case $s=1$, the estimated jet luminosity becomes $ L_{X} \sim \eta\left(\frac{10 r_s}{R_{\rm WD}}\right) \dot{M} c^2 \sim 10^{48} \mathrm{erg/s}$.


\subsection{Late optical flare}
The unbound material $m_{ub}$ will remain on orbits near the MBH and has a significant chance of eventually accreting onto the MBH, producing a TDE-like optical flare.
The delay between the X-ray burst and the optical flare can be estimated as
\begin{align}
\tau \sim 2 \pi \sqrt{\frac{r^3}{GM}} \sim 1\, {\rm day}
\label{eq:tau-gap}
\end{align}
for a MBH of $10^6M_\odot$ and $r=100R_g$.

\subsection{Gravitational Wave}
The strong gravitational interactions between WDs and black holes are also expected to produce gravitational wave emissions \citep{Rosswog:2009,Xuan:2025}.
For an encounter event at distance $D$, the characteristic frequency and strain of the gravitational waves emitted can be estimated as (for $b \ge R_{\rm WD}$)
\begin{eqnarray}
    f_{GW}&\sim&\frac{v_0}{b}\sim 10~\mathrm{Hz} \frac{R_{\rm WD}}{b},\\
    h_{GW}&\sim&\frac{Gm_{\rm WD}v_0^2}{c^4 D}\sim 0.5\times10^{-24}\left(\frac{10^3~\rm Mpc}{D}\right).
\end{eqnarray}

\section{Conclusion}\label{sec:con}

In this work, we have studied the scenario in which a WD undergoes a very close encounter with a sBH. The WD may belong to the stellar population that orbits around the massive black hole at a few hundred gravitational radii, as a result of disk-assisted migration in the previous active phase of the AGN. The relevant sBHs are closely related to those that eventually form EMRIs, as they generally have low angular momentum because of multi-body scatterings in the nuclear star cluster. We estimate that a close encounter between a WD and a sBH may occur once every ten years within $z \le 1$, with higher rate at higher redshifts. Using a set of hydrodynamical simulations, we have quantified the mass accreted onto the sBH and the disrupted, unbound mass that may later accrete onto the MBH, as a function of the initial impact parameter and speed. We find that the amount of bound material monotonically increases with the impact parameter, ranging between $0-20\%$ for $v_0/c \in (0.05,0,2)$ for head-one collisions. On the contrary, the amount of mass captured by the sBH decreases with the impact parameter. It also strongly depends on the initial speed $v_0$: for head-on collisions, with $v_0/c =0.2$ the accretion percentage is almost $100\%$ and for $v_0/c=0.05$ the accretion percentage is nearly zero.
The mass of the unbound material ranges from nearly zero to $\ge 50\%$, reaching the peak value when the impact parameter is comparable to the WD's radius. We expect such events to produce bright X-ray bursts that last $\sim 10$ seconds and low-frequency gravitational wave radiations of $10$~Hz, possibly followed by a weak tidal disruption event after days or months depending on the mass of the MBH.

Einstein Probe is a wide-field X-ray time domain space telescope \citep{Yuan:2018} capable of observing X-ray transients like micro-TDEs. The newly discovered EP240408a with a duration of $12$~s has prompted different follow-up observations. The properties are inconsistent with any known transients so far, including a jetted tidal disruption event, a gamma-ray burst, an X-ray binary and a fast blue optical transient \citep{Zhang:2025}. However, it could be a micro-TDE candidate, if a weak TDE is observed within around one day to months according to the estimation of Eq.~\eqref{eq:tau-gap}.

The characteristic frequency lies at the low-frequency end of the LIGO band, and the strain is orders of magnitude smaller than that from binary black hole or neutron star systems. The characteristic strain of third-generation gravitational wave detectors, such as the Cosmic Explorer \citep{Srivastava_2022}, is around $10^{-24}$ at $10$Hz. For cases with impact radius $b$ significantly larger than $R_{\rm WD}$, the disruption is negligible and the characteristic frequency is lower. However, the rate increases by a factor of $b^2/R^2_{\rm WD}$. Deci-hertz gravitational wave detectors, such as DECIGO \citep{Kawamura:2008zza}, may be suitable for the detection.

Future observational confirmation of such events would not only provide a natural mechanism for producing micro-TDE events, but also serve a powerful probe of the stellar distribution in galactic centers, complementing QPEs and future gravitational wave measurements of EMRIs.

\bibliographystyle{aasjournal}
\bibliography{references}
\end{document}